\documentclass[%
aps, prb,two-column,
 amsmath,amssymb,
 reprint,%
]{revtex4-2}

\usepackage{comment}
\usepackage{braket}
\usepackage{subfigure}
\usepackage{graphicx}
\usepackage{amsmath}
\usepackage{url}
\usepackage{color}
\usepackage{verbatim}
\usepackage{gensymb}
\usepackage[utf8]{inputenc}
\usepackage{multirow}
\usepackage[english]{babel}
\usepackage{amsfonts}
\usepackage{amssymb}
\usepackage{enumerate}
\usepackage{sidecap}
\usepackage{bbold}
\usepackage[normalem]{ulem}
\usepackage{mathtools}
\usepackage[usenames, dvipsnames]{xcolor}
\usepackage{natbib}
\usepackage[colorlinks=true,plainpages=true,citecolor=blue,linkcolor=blue]{hyperref}
\usepackage{cancel}
\usepackage{enumitem}
{\begingroup\color{#1}} 
{\endgroup} 

\makeatletter
\def\@email#1#2{%
 \endgroup
 \patchcmd{\titleblock@produce}
  {\frontmatter@RRAPformat}
  {\frontmatter@RRAPformat{\produce@RRAP{*#1\href{mailto:#2}{#2}}}\frontmatter@RRAPformat}
  {}{}
}%
\makeatother
\newcommand{\Sigmaxc}{\Sigma_{\mathrm{xc}}}
\newcommand{\E}{\text{E}_0}
\newcommand{\Ek}{\mathrm{E}_{\mathrm{kin}}}

\newcommand{\Exc}{\text{E}_{\mathrm{xc}}}
\newcommand{\Wrpa}{W^{\mathrm{RPA}} }
\newcommand{\vext}{v_{\mathrm{ext}}}

\newcommand{\vH}{v_{\mathrm{H}}}
\newcommand{\vks}{v_{\mathrm{KS}}}
\newcommand{\vxc}{v_{\mathrm{xc}}}
\newcommand{\fxc}{f_{\mathrm{xc}}}
\newcommand{\vcl}{v_{\mathrm{cl}}}
\newcommand{\vcoul}{v_{\mathrm{c}}}
\newcommand{\Wb}{\mathcal{W}}
\newcommand{\vb}{\nu}
\newcommand{\Gammab}{\gamma}
\newcommand{\Pb}{\mathcal{P}}

\newcommand{\Vb}{\mathcal{V}}
\newcommand{\chib}{\mathcal{X}}
\newcommand{\nb}{\rho}
\newcommand{\ehks}{\epsilon_1^{\text{\tiny KS}} }
\newcommand{\elks}{\epsilon_2^{\text{\tiny KS}} }

\newcommand{\Gb}{\mathcal{G}}
\newcommand{\GH}{G_{\mathrm{H} }}

\newcommand{\1}{^{(1)}}
\newcommand{\2}{^{(2)}}
\newcommand{\bb}{\bar }
\newcommand{\se}{^{\rm sc}}

\newcommand{\Gks}{G_{\mathrm{KS}}}

\newcommand{\luc}[1]{\textcolor{magenta}{#1}}

\newcommand{\secord}{second-order}
\newcommand{\firstord}{first-order}

\begin{document}

\title{Consistent perturbation expansions in screened interactions}
\author{Abdallah Arthur El Sahili}
\affiliation{LSI, CNRS, CEA/DRF/IRAMIS, \'Ecole Polytechnique, Institut Polytechnique de Paris, F-91120 Palaiseau, France}
\affiliation{European Theoretical Spectroscopy Facility (ETSF)}

\author{Francesco Sottile}
\affiliation{LSI, CNRS, CEA/DRF/IRAMIS, \'Ecole Polytechnique, Institut Polytechnique de Paris, F-91120 Palaiseau, France}
\affiliation{European Theoretical Spectroscopy Facility (ETSF)}

\author{Lucia Reining}
\affiliation{LSI, CNRS, CEA/DRF/IRAMIS, \'Ecole Polytechnique, Institut Polytechnique de Paris, F-91120 Palaiseau, France}
\affiliation{European Theoretical Spectroscopy Facility (ETSF)}

\date{\today}

\begin{abstract}
Many-body perturbation theory is a powerful framework to calculate spectroscopic quantities, as well as the total energy. However, the field suffers from the fact that different flavors of approximations lead to a wide range of results. We present several in principle exact expansions and discuss how to use them consistently. We concentrate on the first two orders in the respective screened Coulomb interaction. Of special interest is 
 a consistent expansion around the Kohn-Sham system.
We illustrate the performance of the various choices using the symmetric Hubbard dimer over a wide range of correlation strength. In particular, the expansion based on exact Kohn-Sham ingredients stands out as a promising route towards good energies and good spectra. 
 
\end{abstract}
\keywords{MBPT, GW, \secord, spectral function, total energy}
\maketitle

\section{Introduction}
The many-body problem is one of the outstanding challenges of condensed matter physics and chemistry. Its exact solution cannot be found analytically or numerically, except for a very small number of particles \cite{Kohn1999}. The complication arises both from the quantum nature of the constituents of matter, and from the fact that all particles are linked by the Coulomb interaction. It is the latter that makes the problem untractable. One widely used class of approaches is therefore  
Many-Body Perturbation Theory (MBPT)\cite{Martin_Reining_Ceperley_2016}, where Green's functions or their related effective potentials, i.e., the self-energies that are the kernels in Dyson-like equations, are expanded in terms of the interaction. It was recognized early on that expansions in the bare interaction may show unsatisfactory behavior, and one has to find ways to renormalize the interaction \cite{kadanoffbaym}. This typically introduces new Dyson equations for these effective interactions. 
The best example is the use of a screened Coulomb interaction $W$ instead of the bare $v_c$, which makes the effective interaction short ranged in a metal and strongly reduced in semiconductors. Screening is crucial especially when one is interested in extended systems. Early on, $W$ was calculated in the homogeneous electron gas by neglecting exchange-correlation effects in the response, which leads to the Random Phase Approximation (RPA) \cite{bohmpinesI,pinesbohmII,bohmpinesIII}. It was found that it is mandatory to solve the Dyson-like equation for the response function, because an order-by-order expansion of the RPA energy in the homogeneous electron gas diverges \cite{Gell-Mann1957}. 

Indeed, more generally, MBPT is not a well defined and well behaved textbook perturbation expansion. Even if one simply examines the structure of the equations as if all ingredients were scalars, it is clear that one encounters finite convergence radia. For example, the solution of a Dyson equation is of the form $G=(1-G_0\Sigma)^{-1}G_0$, where $G$ is the quantity of interest, e.g., the one-body Green's function, $G_0$ is its non-interacting counterpart, and the self-energy $\Sigma$ contains all interaction effects. This means that an order-by-order expansion of the Dyson equations corresponds to expanding $1/(1-x)$, which has a converence radius of 1. A scalar model of $G$ itself as a function of the interaction suggests asymptotic convergence \cite{Lani2012}, which means that for a given interaction strength one is limited to a certain number of orders. This also implies that re-ordering the series might change the results. The situation is, of course, even more complicated in reality, since one is working with functions that have poles. Still, one can understand a few guidelines. First, one should aim to expand in a renormalized interaction that is as small as possible over the range of interest. Second, Dyson equations are efficient ways to resum at least a part of a series, and it is sometimes, but not always, as we will see, a good strategy to approximate the kernel of a Dyson equation instead of approximating directly its solution. This is the reason why one often expands the self-energy, and not the Green's function itself. Third, even if one leaves aside the fact that higher order terms are usually more clumsy to calculate, one should head for an expansion where already the first few orders, ideally one or two, give sufficiently good results, since there is no guaranty that higher orders would lead to improvement. 

One of the most widely used approximations in MBPT in condensed matter is Hedin's $GW$ approximation \cite{Hedin1965}, which is a \firstord{} expansion of the exchange-correlation (xc) contribution to the self-energy, $\Sigmaxc$, in terms of a screened Coulomb interaction $W$. It is widely recognized as the standard method to calculate the band structure of materials \cite{Strinati1980,Hybertsen1985,Godby1987,Godby1988,Blase1995,Schilfgaarde2006,Kotani2007} and the energy differences in atoms and molecules \cite{Shirley1993,Grossman2001,Rostgaard2010,Blase2011,Bruneval2012,Thomas2012,Bruneval2013,VanSetten2013,Koval2014,VanSetten2015,Knight2016,Bruneval2021}. However, there is no unique choice for what $W$ exactly is, which is one of the main reasons for the fact that $GW$ results of different authors, while they often agree qualitatively, still show quantitative discrepancies. Another reason for these differences is the second ingredient of the self-energy, which is the one-body Green's function $G$ used in input to build the self-energy, and which is also chosen according to various recipes. Indeed, strictly speaking MBPT is an expansion in a quantity $\Gb\Gb \Wb$, where $\Gb$ is some (the exact or an approximate) one-body Green's function and $\Wb$ is some effective interaction, so one has to choose both $\Gb$ and $\Wb$.  

In many cases, the choices for $\Gb$ and $\Wb$ are made based on intuition or for purely pragmatic reasons. The input Green's function $\Gb$ is often obtained from a Kohn-Sham (KS) \cite{KohnSham1965} or generalized KS \cite{Seidl1996} calculation of Density Functional Theory (DFT) \cite{Hohenberg1964}, using different density functionals. These functionals are generally not diagrammatic, which makes it difficult to estimate their role in the subsequent MBPT. A Hartree-Fock starting point can also be found, especially for atoms and molecules \cite{Bruneval2013}. Sometimes $\Gb$ is determined in a partially or, more rarely, in a fully self-consistent way, which adds well-defined diagrams and which has the benefit that it may correct the density and one-body reduced density matrix with respect to (approximate or, in the case of the density matrix, also exact) KS calculations \cite{Bruneval2006,Bruneval2019,Bruneval2021b}. Especially for systems with localized electrons, self-consistency within the quasi-particle (QP) approximation has led to strong improvements \cite{Faleev2004,Vanschilfgaarde2006,Gatti2007}. However, because some diagrams have a tendency to cancel \cite{Martin_Reining_Ceperley_2016}, the results are not always guaranteed to improve with self-consistency. In particular, it is not clear whether one should also calculate $\Wb$ using the self-consistent $\Gb$ \footnote{For example, an RPA screened interaction built with dressed Green's functions does not fulfill the f-sum rule}\cite{Holm1998,Shishkin2007,Schindlmayr1998,Wen2024}. On the other hand, there is no proof that the best choice for $\Wb$ in the $GW$ approximation should be the exact measurable screened Coulomb interaction. Often, this does not seem to be the case \cite{Lewis2019,Vlcek2019}.   

The present work aims at putting the various choices on a common footing in order to contribute to clarifying these issues, starting from the screened interaction and including the choice of the Green's function used in the perturbation series. All approximations are derived from the same initial expression in terms of Schwinger's functional derivative approach \cite{Martin1959,Schwinger1951,Baym1962}. We derive the rigorous perturbation expansion up to second order for the general case. We then examine the particular cases of an RPA screened Coulomb interaction $\Wrpa$, of the exact measurable, i.e., test charge-test charge $W$, and of a test charge-test electron screened Coulomb interaction $\tilde W$.
We put special emphasis on the fact that the various contributions to the total energy as well as the spectral functions do not necessarily behave in the same way, and we highlight the link between the choice of $\Wb$ and the choice of $\Gb$.

There are different arguments in favor of the three choices that we examine. 
The use of RPA screening for $\Wb$ is common in calculations of real materials  \cite{Martin_Reining_Ceperley_2016,Benites2024}; it is considered to be consistent with the $GW$ approximation \cite{Hedin1965}. The use of the exact $W$, on the other hand, is appealing, because $W$ contains the exact excitation energies, and one would intuitively expect that this is important, for example, for satellites in photoemission spectra. The third option that we examine, namely, the test charge-test electron screened interaction $\tilde W$, is also supported by physical arguments:
whereas the test charge-test charge screening in $W$ 
reflects the fact that propagating particles induce a change in the Hartree potential, 
the test charge-test electron screening in $\tilde W$ in addition accounts for an induced xc potential \cite{Romaniello2009,Martin_Reining_Ceperley_2016}. In a full description, this induced (non-local) xc potential would be due to a variation of the (unknown) self-energy, and taking it into account would make 
 $\Sigmaxc=iG\tilde W$ exact, if $G$ is the exact Green's function (which in this case would be integrated with $\tilde W$ that depends on three space, spin and time arguments \cite{Reining2018}). As an approximation, this complicated $\tilde W$ may be calculated from an approximate Bethe-Salpeter equation, i.e., from a Dyson-like equation whose interaction kernel contains the variation of an approximate $\Sigmaxc$ with respect to $G$ \cite{Kuwahara2016,Maggio2017,Rohlfing2023,Foerster2025,Vlcek2019,Kutepov2016}. Still, these calculations remain involved, although recent computational advances open new possibilities \cite{Vlcek2019}.
Instead, $\tilde W$ has more often been approximated by taking the induced local exchange-correlation potential from Time-Dependent Density Functional Theory (TDDFT) \cite{RungeGross1984}. This reduces the complexity enormously, because the resulting $\tilde W$ depends only on two space-time arguments. Even with the exact TDDFT ingredients, however, now $\Sigmaxc=iG\tilde W$ is an approximation that impacts quantities derived from it, except for the total xc energy, which remains exact\cite{elsahili2023total}. Such an incorporation of mean-field corrections into many-body perturbation theory was proposed by several authors
\cite{Streitenberger1984,Mahan1989,DelSole1994,Hindgren1997,Bruneval2005,Tiago2006,Hellgren2018b}. Usually, additional approximations are made to the resulting $\Sigmaxc=iG\tilde W$, since also the exact TDDFT ingredients are unknown. Nevertheless, some encouraging results were obtained, e.g., in Ref.s \cite{Chen2015,Maggio2017,Olsen2019,Tal2021,Abdallah2024}.

Studies beyond a \firstord{} expansion in some effective $\Wb$ are less frequent. Besides works based on diagrammatic Monte Carlo methods \cite{Vanhoucke2010}, they are mostly (though not always, see, e.g., \cite{Mejuto2022,Cunningham2024}) limited to second order, often with mixed results \cite{Minnhagen1974,Bobbert1994,Gruneis2014,Hinuma2014,Ren2015,Kutepov2016,Kutepov2017,Pavlyukh2016,Bruneval2024,Schindlmayr1998,Wang2021,Gruneis2014,Foerster2022,Yoneyama2024}. One major problem is for example the appearance of negative spectral functions, which has to be overcome by modified diagrammatic expansions \cite{Pavlyukh2016,Stefanucci2014,Stefanucci2016,Uimonen2015,Pavlyukh2020}. It is difficult to draw conclusions, not least because of the interplay of different aspects such as self-consistency, dynamical effects and vertex corrections \cite{Kutepov2022}; indeed, cancellations between these aspects were mentioned early on \cite{Ummels1998}. With the large spread of ingredients, approximations and observables of interest, a clear picture still has to emerge. 

The aim of the present work is to define what we may call a consistent approximation for the various choices of ingredients, and to explore whether, and if yes why, certain choices are to be preferred. We focus on a well defined and self-contained way to derive the straightforward diagrammatic series, and we do not touch on questions such as self-consistency, the use of the Bethe-Salpeter equation, or other kinds of approximations that add selected diagrams \cite{Hummel2019,Ren2013,Ren2015,Vacondio2022}, which are interesting but which would blur our focus. 
 Some important points will be illustrated with results for the symmetric Hubbard dimer, where approximations can be compared to the exact solution.

The paper is organized as follows: we start by providing the fundamental equations in Sec.~\ref{Theoretical back}. In Sec.~\ref{theoretical-dev-sec}, we show how MBPT can be, in principle, expressed in terms of many different Green's functions and screened interactions. In Sec.~\ref{sec-dressed}, particular cases are examined. We illustrate our findings using an exactly solvable model, namely the symmetric Hubbard dimer at half filling, in Sec.~\ref{sec:results}, where we show comparisons between the different expansions by calculating ground-state total energies, QP energies, and spectral functions.
Finally, we conclude the article in Sec.~\ref{sec:conclusions} and provide analytical formulas in the Appendices. 

\section{Theoretical background \label{Theoretical back}}
In the present work we concentrate on the calculation of the equilibrium one-body Green's function $G$ at zero temperature. It is obtained from the Dyson equation that reads schematically 

\begin{equation}
    G^{-1}(1,2)= G_0^{-1}(1,2) - \delta(1,2)\vcl(1) - \Sigmaxc(1,2)\,,
    \label{eq:dyson-inverse}
\end{equation}
or, with the appropriate boundary conditions, \cite{book:vL}
\begin{multline}
 G(1,2) = G_0(1,2) \\ +  \,G_0(1,\bb 3)\bigg( \vcl(\bar 3)\delta(\bar3,\bar4) +\Sigmaxc(\bb 3,\bb 4)\bigg)G(\bb 4,2)\,,
\end{multline}
with the exact xc self-energy 
\begin{equation}
\Sigmaxc(1,2) = -i  G(1,\bb 4) \vcoul(\bb 3,1^+) \frac{\delta G^{-1}(\bb 4,2) }{\delta \vext(\bb 3)}\,.
\label{xc self-energy}
\end{equation}
Here, 
$1 \equiv (x_1,t_1) \equiv (r_1,\sigma_1,t_1)$ stands for position, spin and time, with $t^+ = t + \eta$ where $\eta \rightarrow 0^+$. Arguments with a bar are integrated over, as in $\bb 1 \rightarrow \int dx_1 dt_1$. The local external potential $\vext(\bb 3)$ is to be understood as the equal time limit $\vext(\bb 3^{+},\bb 3)$, and $\vcl\equiv\vext+\vH$ is the total classical potential, where the Hartree potential  $v_{\rm H}(3,4)=\delta(3-4)\int d{\bf r}\dfrac{n({\bf r})}{|{\bf r}_3-{\bf r}|}$ is calculated with the exact density $n$, and $\vcoul(1,2) = \delta(t_1-t_2) \dfrac{1}{|r_1-r_2|}$ is the bare Coulomb interaction. The external potential $v_{\rm ext}$ is a perturbation to the system potential contained in $G_0$ that is set to zero once the derivatives are taken. When the Hartree and xc contributions that appear in \eqref{eq:dyson-inverse} are neglected in the functional derivative in \eqref{xc self-energy}, the latter yields $-\delta(\bar 3,2)\delta(\bar 4,\bar 3)$ and the Hartree-Fock approximation is obtained. Beyond Hartree-Fock,
\begin{equation}
    \Gamma(4,2;3)\equiv - \frac{\delta G^{-1}(4,2) }{\delta \vext(3)}
\end{equation}
contains additional terms called vertex corrections. 

From the time-ordered $G$, spectral functions are obtained by evaluating
\begin{equation}
 A(x,x,\omega) = \frac{{\rm sgn}\,(\omega-\mu)}{\pi} {\rm Im}\, G(x,x;\omega) \,, 
 \label{spf in terms of the 1-GF}
\end{equation}
and the ground-state total energy $\E$ can be expressed in terms of the GF and the self-energy as \cite{Galitskii1950} 
\begin{multline}
\E =
- i\lim_{x_1'\to x_1} \int dx_1\, \bigg[ h_0(x_1)  + \frac{\vH (x_1)}{2} \bigg]G(x_1,x_1';t_1,t_1^+)\\
- \frac{i}{2}  \lim_{t_2 \to t^{++}_1} \!\! \int \!\! dx_1d(3) \Sigmaxc(x_1,x_3;t_1,t_3)G(x_3,x_1;t_3,t_2), 
\label{GM in self-energy}
\end{multline}
where $h_0(x_1) = - \frac{\nabla^2_{r_1}}{2} + \vext(x_1)$ leads to the kinetic and external potential energies, and where the last line is by definition the  xc potential energy $\Exc$.

Iteration of  \eqref{xc self-energy} starting from some guess for $G$ creates the diagrammatic expansions of many-body perturbation theory in terms of the bare Coulomb interaction for the self-energy, and hence for all the quantities derived from it.  

In order to overcome the shortcomings of the bare interaction, Hedin used the total classical potential $v_{\rm cl}$ to screen the interaction \cite{Hedin1965}. Screening naturally emerges when $v_{\rm cl}$ is used in a chain rule, 
\begin{align}
\label{eq:chain-1}
 \Sigmaxc(1,2) &= -i  G(1,\bb 4) \vcoul(\bb 3,1^+) \frac{ \delta G^{-1}(\bb 4,2)  }{ \delta \vcl(\bb 5)  }  \frac{\delta \vcl(\bb 5)}{ \delta \vext(\bb 3)}\\
 &\equiv -i  G(1,\bb 4) W(\bb 5,1^+)  \frac{ \delta G^{-1}(\bb 4,2)  }{ \delta \vcl(\bb 5)  }\,,
 \label{eq:chain-cl}
\end{align}
which defines the screened interaction $W$, and which introduces the irreducible vertex 
\begin{equation}
    \Gamma^{\rm irr}(4,2;5)\equiv -\frac{ \delta G^{-1}(4,2)  }{ \delta \vcl(5)  }.
\end{equation}
Now, it is easy to keep the full $\vcl$ in $\delta G^{-1}/\delta \vcl$, neglecting only $\Sigmaxc$, which yields the $GW$ approximation 
\begin{equation}
    \Sigmaxc(1,2) \approx i G(1,2)W(2,1^+)\,.
\end{equation}
Both Hartree-Fock and $GW$ are \firstord{} expressions in their respective effective interactions, but one may expect $GW$ to perform better than Hartree-Fock, simply because less terms are neglected in the functional derivative of $G^{-1}$. Indeed, if one would insert the full $v^{\rm tot}\equiv v_{\rm ext}+\vH+\Sigmaxc$ in the chain rule, one could keep all terms, find $\Gamma^{\rm tot}=1$, and the resulting \firstord{} expression 
\begin{equation}
    \Sigmaxc(1,2)= i  G(1,\bb 4) W^{\rm tot}(2,\bb 4;1^+)
    \label{eq:sigmaxctot}
\end{equation}
would be exact \cite{Reining2018} - however, at the price of a 3-arguments and difficult to obtain $W^{\rm tot}(2,4;1^+)$,
\begin{equation}
    W^{\rm tot}(2,4;1^+)\equiv v_c(\bb 3,1^+)\frac{\delta v^{\rm tot}(4,2)}{\delta v_{\rm ext}(\bb 3)}\,.
\end{equation}
In other words, various series can be created, and the choice to be made is a trade-off involving the importance of higher orders, the difficulty to obtain the low orders, and the convergence properties. This will be investigated in the following section. Moreover, note that up to here only the exact, and therefore, unknown, Green's function appears in all expressions. How to handle this fact is a second important point in MBPT, and it will also be discussed below.
\section{Dressed ingredients: general formalism \label{theoretical-dev-sec} }

In this section, we will examine exact and perturbative expressions for the self-energy, for the total xc energy, and for the Green's function and related properties.  

\subsection{Generation of equations involving screened interactions }
\label{freedom in FD}
In the following, we generalize \eqref{eq:chain-1} to

\begin{align}
 \Sigmaxc(1,2) &= -i  G(1,\bb 4) \vcoul(\bb 3,1^+) \frac{ \delta G^{-1}(\bb 4,2)  }{ \delta \vb(\bb 5)  }  \frac{\delta \vb(\bb 5)}{ \delta \vext(\bb 3)}\nonumber\\
 &= -i  G(1,\bb 4) \Wb(\bb 5,1^+)  \frac{ \delta G^{-1}(\bb 4,2)  }{ \delta \vb(\bb 5)  }\,.
 \label{eq:sigmaxc-vbar}
\end{align}
Here, 
\begin{equation}
    \Wb (5,1^+) \equiv \vcoul(\bb 3,1^+)\frac{\delta \vb(5)}{ \delta \vext(\bb 3)}
\end{equation}
is an effective screened Coulomb interaction. The use of this chain rule is clearly justified when 
the effective potential $\vb$ is an invertible functional of $\vext$, i.e., when there is a one-to-one relation between $\vb$ and $\vext$. This point is briefly discussed for all choices of $\vb$ considered here in App.~\ref{sec:app-chain}. Note that  we take into account only local potentials, in order to keep a reasonably simple effective interaction.

The choice $\vb=\vcl$ is one possibility out of many. Using any allowed $\vb$, the generalization of Hedin's equations, besides the Dyson equation, reads
\begin{eqnarray}
    \Sigmaxc(1,2)&=& iG(1,\bar 4) \Wb(\bar 5,1^+) \Gammab(\bar 4,2;\bar 5)\\
    \label{eq:W}
    \Wb(2,1^+) &\equiv& \frac{\delta \vb(2)}{\delta \vext(\bar 3)}v_c(\bar 3,1^+)\\
    \label{eq:W-chi}
    &=& \vcoul(2,1^+) + \Vb(2,\bar 4)\,\chib(\bar 4,\bar 3)\vcoul(\bar 3,1^+)\\
    \label{eq:W-P}
    &=& \vcoul(2,1^+) + \Vb(2,\bar 4)\, \Pb(\bar 4,\bar 3) \Wb(\bar 3,1^+)\\ 
    \Gammab(4,2;5) &=& -\frac{ \delta G^{-1}(4,2)  }{ \delta \vb(5)  }\\    \label{eq:chi-bar}
    \chib(1,2) &=& \frac{\delta \nb(1)}{\delta \vext(2)}\\
    \label{eq:p-bar}
    \Pb(1,2) &=& \frac{\delta \nb(1)}{\delta \vb(2)}\,\\
     \label{eq:V-bar}
    \Vb(1,2) &=& \frac{\delta (\vb(1)-\vext(1))}{\delta \nb(2)}\,,
\end{eqnarray}
where $\nb$ is an appropriately chosen density that we use in a chain rule for the derivation of \eqref{eq:W-chi}, and, similarly, for \eqref{eq:W-P}: 
\begin{equation}
    \frac{\delta \vb(2)}{\delta \vext(\bar 3)} = \delta(2,3) + \frac{\delta (\vb(2)-\vext(2))}{\delta \nb(\bb4)}\frac{\delta \nb(\bb 4)}{\delta \vext(\bar 3)}\,.
    \label{eq:chain-barn}
\end{equation}
Note that this is not an approximation, even if $\rho$ is not the exact density, as long as its use in a chain rule is admitted (see \ref{sec:app-chain}). One will in general opt for a density $\nb$ in terms of which $\vb$ is known explicitly. In particular, we will use $\nb=n$, the exact density, when $\vb$ is a known functional of the latter, or $\nb$ will be 
the density that is obtained from the diagonal of a $\Gb$ that is solution of the Dyson equation with kernel $\vb$, 
\begin{equation}
    \Gb = G_0 + G_0 \vb \Gb\,.
    \label{eq:gbar}
\end{equation}
In all cases, it will be important to specify the choice that is made; it enters the definition of the generalized reducible and irreducible polarizabilities \eqref{eq:chi-bar} and \eqref{eq:p-bar}. It also leads to an effective ``bare'' interaction $\Vb$ that equals the bare Coulomb interaction only when $\vb$ is the total classical potential where the Hartree potential is calculated from the chosen $\nb$, otherwise \eqref{eq:V-bar} has to be used.

\subsection{Perturbation theory in the screened interaction}
A perturbation series is obtained by iterating \eqref{eq:sigmaxc-vbar}. 
To this purpose, we write \eqref{eq:dyson-inverse} as
\begin{equation}
    G^{-1} = G_0^{-1} - \vb - \bigg( \Sigmaxc + \vcl - \vb \bigg)\,
    \label{eq:dyson-vb}
\end{equation}
and neglect the terms in parenthesis at the first or later iterations of \eqref{eq:sigmaxc-vbar}. In this way, the \firstord{} self-energy reads 
\begin{equation}
    \Sigmaxc\1(1,2)=iG(1,2)\Wb(2,1^+) \,,
    \label{eq:sigmaxc-1b}
\end{equation}
where $\Wb$ is defined in \eqref{eq:W-chi} and the following equations, and the vertex is approximated $\Gammab(4,2;5)\1 \approx\delta(4,5)\delta(2,5)$. 
For the \secord{} contribution, one has to include the derivative of the first-order expression of the parenthesis in \eqref{eq:dyson-vb}. This yields
\begin{align}
\Sigmaxc\2(1,2) =&  i \, G(1,\bb 4) \Wb(\bb 5,1^+) \nonumber \\ 
&\times \frac{  \delta \bigg(\Sigmaxc\1(\bb 4,2)+\delta(\bb 4,2)(\vcl(2)-\vb(2))\1 \bigg)}{\delta \vb(\bb 5)} \label{eq:sxc-2-step1} \\
=&  i \,G(1,\bb 4) \Wb(\bb 5,1^+) \bigg(iG(\bb 4,\bb 5) G(\bb 5,2)  \Wb(2,\bb 4^+)\nonumber \\ & + \delta(\bb 4,2) \Big [\frac{\delta(\vcl(2)-\vb(2))}{\delta \vb(\bb 5)}\Big ]\1\bigg)\,.
\label{eq:sigmaxc-2-g}
\end{align}

Note that the equal signs in \eqref{eq:sxc-2-step1} and \eqref{eq:sigmaxc-2-g} are to be interpreted as equal to \secord{}; in particular, from \eqref{eq:sxc-2-step1} to \eqref{eq:sigmaxc-2-g} we have used that 
 $\frac{\delta \Wb}{\delta \vb} = - \Wb \frac{\delta \Pb}{\delta \vb} \Wb$ is at least of \secord{} in $\Wb$ and, since integrated with $G\Wb$, is therefore to be neglected here. Moreover, the zero-order expression $\frac{\delta G}{ \delta \vb }\approx GG$ is to be used. 
Also note that the functional derivative contribution in the last line disappears when $\vb=\vcl$, but otherwise one has to take this extra term into account, while paying attention to include only its \firstord{} contribution, as in the case of the derivative of the self-energy. This will be important later, when we use for $\vb$ the total KS potential.

\subsection{Perturbation theory in the screened interaction and in an effective Green's function}
\label{subsec:gf}

The exact Green's function that appears in the above expressions is not known, and a strategy is needed also for this issue. Two well defined possibilities are to either calculate $G$ self-consistently from the Dyson equation once the self-energy expression is defined, or to expand also $G$, starting from $\Gb$ defined in \eqref{eq:gbar}. Here, we will follow the second path, which is easier to put into practice since it avoids any self-consistency. Most importantly, it yields exact results at all orders for $\Exc$ in the particular case of $\Gb = \Gks$ the KS Green's function, as we will see in Subsection \ref{sec:pt-ks}. We therefore expand 
$G=\Gb + \Gb (\vcl+\Sigmaxc-\vb)G$ 
to the required order, with the hope to create diagrams that make the perturbation expansion consistent and lead to better results. 

The new \firstord{}{} self-energy is then obtained from  \eqref{eq:sigmaxc-1b} by replacing $G$ with the zero-order $\Gb$, which yields
\begin{equation}
    \Sigmaxc\1 (1,2) = i\Gb(1,2)\Wb(2,1^+) \,.
    \label{eq:sigmaxc-1bb}
\end{equation}
The \secord{} contribution to the self-energy has now the terms in \eqref{eq:sigmaxc-2-g} with $G$ replaced by $\Gb$, plus a term $\Sigmaxc^{\rm sc(2)}$ stemming from the \firstord{}{} expression \eqref{eq:sigmaxc-1b} with $G$ replaced by the first-order contribution to the self-consistent Green's function $G^{(1)}=\Gb\,(i\Gb \Wb+\vcl\1  - \vb\1)\Gb$,  
\begin{multline}
    \Sigmaxc^{\rm sc(2)}(1,2) = i\Gb(1,\bb 3)\,\bigg( i\Gb(\bb3,\bb4)\Wb(\bb4,\bb3^+)  \\ + \left( \vcl\1(\bb3)  - \vb\1(\bb3) \right) \delta(\bb3,\bb4) \bigg) \Gb(\bb 4,2)\Wb(2,1^+)\,.
    \label{eq:sc2}
\end{multline}
Again, note that in general, $\vb$ may contain itself interaction effects. In this case, one has to be careful to make a consistent expansion. This implies that here, only the \firstord{}{} contribution $\vb\1$ has to be taken, instead of the full $\vb$. This will again become more obvious in Subsection \ref{sec:pt-ks}.

\subsection{Determination of the Green's function and derived quantities}

\subsubsection{Green's functions and QP equations}

Once the self-energy has been calculated, there are different ways to obtain the Green's function. One may fully solve the Dyson equation, 

\begin{multline}
  G^{(Dn)}(\omega) = \Gb (\omega) \\ + \Gb (\omega) \Big ( \Sigmaxc^{(sn)}(\omega) - \left( \vb - \vcl \right)^{(sn)} \Big ) G^{(Dn)}(\omega) \,,
  \label{eq:dysonb-full)}
\end{multline}
where $(sn)$ denotes a quantity  that sums orders up to order $(n)$, and where $Dn$ indicates a solution of a Dyson equation with a self-energy kernel $(sn)$. Alternatively, one may expand the Dyson equation
such that the Green's function is calculated consistently to the given order, schematically,
\begin{multline}
G^{(s2)}(\omega) = \Gb (\omega) + \Gb (\omega) \Big ( \Sigmaxc^{(s2)}(\omega) - \left( \vb - \vcl \right)^{(s2)}\Big ) \Gb(\omega)\\
+ \Gb (\omega) \Big ( \Sigmaxc^{(1)}(\omega) - \left( \vb - \vcl \right)^{(1)}\Big )  \Gb (\omega)\\ \times \Big ( \Sigmaxc^{(1)}(\omega) - \left( \vb - \vcl \right)^{(1)}\Big ) \Gb(\omega)\,.
\label{eq:dysonb-ede)}
\end{multline}
Both the full solution of the Dyson equation (FDE) and the expanded Dyson equation (EDE) should converge to the exact result when the expansion is taken to infinity.  However, they will perform very differently when the series is truncated.

From the full solution of the Dyson equation and when everything is supposed to be diagonal in the same basis, one obtains the QP equation
\begin{align}
    \epsilon^{\rm QP}_{\ell} &= \varepsilon_{\mathrm{H},{\ell}} + \Sigma_{\rm xc,{\ell}}(\epsilon_{\ell}^{\rm QP})\nonumber \\
    &= \varepsilon_{\ell} + \Sigma_{\mathrm{xc},{\ell}}(\epsilon^{\rm QP}_{\ell}) - \left( \vb_{\ell} -v_{\mathrm{cl},{\ell}} \right)\,,
    \label{eq:qp-choices}
\end{align}
where $\varepsilon_{\ell}$ is the eigenvalue associated to state $\ell$ of the non-interacting system with potential $\vb$, and all quantities are diagonal matrix elements taken in state $\ell$.   The first expression is obtained from a Dyson equation starting with the exact Hartree Green's function $\GH$ and keeping only the xc self-energy in the kernel, whereas the second, equivalent, one includes $\vb$ in the starting Green's function $\Gb$ and subtracts it from the kernel.
When the imaginary part of the self-energy is small and/or constant, the position of the QP peak in the spectral function $A_{\ell}(\omega)=\frac{1}{\pi}|{\rm Im}\,G_{\ell}(\omega)|$ corresponds to  $\epsilon^{\rm QP}_{\ell}$. 

For the  QP energy associated to a certain order, this leaves two choices: 
\begin{eqnarray}
\label{eq:qp-possibilities-H}
    \epsilon^{{\rm QP} (Dn)}_{\ell} &=& \varepsilon_{{\mathrm{H}},{\ell}} \!+\! \Sigma_{{\rm xc},{\ell}}^{(sn)}(\epsilon_{\ell}^{{\rm QP}(Dn)})\\
    &{\rm or}& \nonumber \\
    &=& \varepsilon_{\ell} + \Sigma_{{\rm xc},{\ell}}^{(sn)}(\epsilon^{{\rm QP}(Dn)}_{\ell})\! -\! \left( \vb_{\ell} \!- \! v_{\mathrm{cl,{\ell}}} \right)^{(sn)}\!\!.
    \label{eq:qp-possibilities}
\end{eqnarray}
For a finite-order approximation, the two equations are no longer equivalent. In order to calculate QP energies, one will then rather use the version where more interaction effects are already contained in the starting $\varepsilon$; in particular, in case of a KS starting point,  $\varepsilon$ is a KS energy that has interaction contributions to all orders, and one will opt for \eqref{eq:qp-possibilities}. However, as we will see later, a detour through \eqref{eq:qp-possibilities-H} will be needed in order to find $\vb^{(sn)}$ in that case. 
 
Finally, one might also think to determine the QP energy from the EDE, by calculating the spectral function and taking the peak positions. However, while the full Dyson equation easily shifts QP energies, even through a static self-energy, this is not true for the EDE, which simply induces multiple poles \cite{Reining2018}. We will therefore not consider this possibility for the QP energies.

\subsubsection{Expansion of the kinetic energy}
The kinetic energy is often calculated with a Green's function that is solution of the Dyson equation. It will be interesting to compare this to a consistent expansion. Since the kinetic energy is linear in the Green's function, we simply need the expansion of $G$ to second order, the EDE \eqref{eq:dysonb-ede)}, which is given with all arguments in the Appendix, see Eq.~\eqref{eq:kinetic-expand}.

The same expansion will also yield order by order the density, and therefore the external potential energy, and it can be used to build a consistent expansion for the Hartree energy. Here, we will not consider density-related effects, which are irrelevant for systems with translational symmetry such as the homogeneous electron gas or the symmetric Hubbard dimer. A separate study of these effects for inhomogeneous systems will be interesting in the future.

\subsubsection{The exchange-correlation energy}

Finally, we can express the total xc energy to first and second orders. 
For this purpose, we have to combine the self-energy and the Green's function, and we will do this in a consistent way, i.e., we use the EDE for the Green's function.
Schematically, from \eqref{GM in self-energy} we have $\Exc\1=-i/2\,\int\Sigmaxc\1\Gb$ and $\Exc\2=-i/2 \int \Big(\Sigmaxc\2\Gb + \Sigmaxc\1 G\1\Big )$. 
This yields the diagrams in Fig.~\ref{Feynman diag o2 self-energy}; the corresponding equations can be found in the Appendix, \eqref{eq:exc-general-1} and \eqref{eq:E-2-general}.

\begin{figure}
    \centering
        \includegraphics[width=\linewidth]{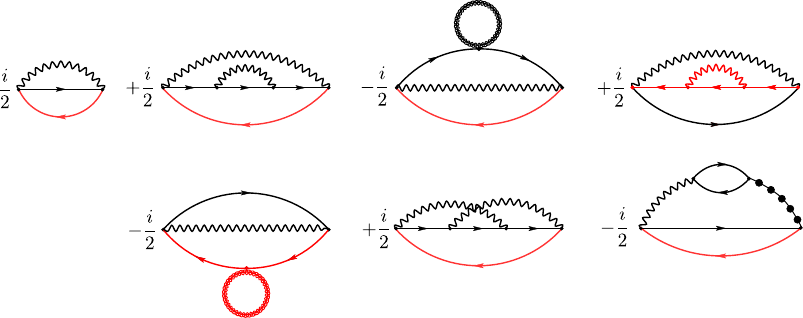}
    \caption{Second-order expansion of the self-energy and of the xc energy. Self-energy in black, without factors $i/2$. The xc energy includes also the factors $i/2$ and all red symbols. A straight arrow indicates $\Gb$, the wiggly line is $i\Wb$. The circle stands for $\vb\1-\vcl\1$, and the line with filled circles is $i$ times its functional derivative with respect to the density.
    The first diagram contains the first-order contribution to the xc self-energy \eqref{eq:sigmaxc-1bb}. The \secord{} contributions to the xc self-energy are the self-consistency diagrams \eqref{eq:sc2} (contained in the next two diagrams), and \eqref{eq:sigmaxc-2-g} with $G\to \mathcal{G}$ (contained in the last two diagrams). The two remaining \secord{} diagrams give an extra contribution only to $\Exc$: they contain again the \firstord{} contribution to the self-energy, supplemented by the \firstord{} contribution of the Green's function that closes $\Exc$ (in red).  }
    \label{Feynman diag o2 self-energy}
\end{figure}

Here, the first-order contribution to the xc self-energy is \eqref{eq:sigmaxc-1bb}. The first two diagrams in the \secord{} are self-consistency contributions to $G$ in the \firstord{}{} self-energy expression, \eqref{eq:sc2}. The next two diagrams stem from self-consistency contributions to the Green's function closing the GM expression, i.e, $-i/2\int \Sigmaxc^{(1)} G^{(1)} $,  and the last two diagrams are \eqref{eq:sigmaxc-2-g} with $G\to \Gb$. Note that in the calculation of $\Exc$ the first and third \secord{} diagrams are equal and add up; the same holds for the second and fourth contribution.

\section{Dressed ingredients: particular cases \label{sec-dressed} }

In the following, we will examine particular choices for $\vb$. In all cases, the expansion is in principle exact, as long as there is no convergence problem. The convergence properties, and in particular, the quality of the first and \secord{} results, may depend crucially on the choice of $\vb$. As we will see, some conclusions can already be drawn from the general formula; others will depend on the parameter range of the application.

\subsection{The total classical potential}

Closest to Hedin's derivation \cite{Hedin1965}, a relatively simple choice is to screen the bare potential by the Hartree contribution. In this case,   
$\vb = \vcl = \vext + \vH$, where 
\begin{equation}
    \vH({\bf r})= \int d{\bf r}'\,\frac{n({\bf r}')}{|{\bf r}-{\bf r}'|}
\end{equation}
with $n({\bf r}')$ the exact charge density. In this case, one will choose $\nb=n$, the exact charge density, in \eqref{eq:W} to \eqref{eq:V-bar}, and $\Vb = v_c$, the bare Coulomb interaction. Moreover, $\chib=\chi$ and $\Pb = P$, the exact reducible and irreducible polarizabilities, respectively, which leads to $\Wb=W$, the exact screened Coulomb interaction. Here, the screening is defined to be the test-charge test-charge screening as measured in experiments with classical perturbations and probes. Following subsection \ref{subsec:gf}, we will use $\Gb=\GH$, the Hartree Green's function obtained from a Dyson equation with the total classical potential of the exact charge density. 
This yields the diagrams in Fig.~\ref{Feynman diag o2 self-energy} with  $\Gb\to \GH$ and $\Wb\to W$. The total classical potential drops out, so the second, fourth, and last \secord{} diagrams do not contribute. The expressions are detailed in Eqs.~\eqref{eq:S-H}, \eqref{eq:exc-H-1}, and \eqref{eq:E-2-H} in the Appendix.

\subsection{The self-consistent total classical potential}
The $GW$ approximation is often seen in conjunction with the RPA for $W$, which is itself closely linked to time-dependent Hartree theory. However, 
in the previous subsection where $\vb$ was the Hartree potential, the exact screened test charge-test charge interaction appeared. In order to make the RPA screening a natural choice, we will work in the following with the self-consistent Hartree potential
$\vb=\vext+v_{\rm H}\se$, where
\begin{equation}
    \vH\se({\bf r})= \int d{\bf r}'\,\frac{n_{\rm H}\se({\bf r}')}{|{\bf r}-{\bf r}'|}\,,
\end{equation}
with $n_{\mathrm{H} }\se $ the density obtained self-consistently from the Hartree equation, i.e., from $\GH\se$ solution of  $\GH\se=G_0 + G_0 \vH\se \GH\se$.
In this case, we choose $\nb=n_{\rm H}\se$ in \eqref{eq:chain-barn}, which yields 
\begin{equation}
    \chib=\chi^{\mathrm{RPA}}\equiv \frac{\delta n_{{\mathrm{H}}}\se}{\delta \vext}= \chi_{0{\mathrm{H}}}\se + \chi_{0{\mathrm{H}}}\se v_c \chi^{\mathrm{RPA} }\,,
\end{equation}
where $\chi_{0\mathrm{H}}\se\equiv -i\GH\se \GH\se $ is the independent-particle polarizability built with the self-consistent Hartree Green's function. $\chi^{\rm RPA}$ is then used in \eqref{eq:W-chi} to calculate the RPA screened Coulomb interaction $\Wrpa$, which is now our $\Wb$. The natural choice for $\Gb$ is $\Gb \to \GH\se$. With this, we get all the terms of 
Fig.~\ref{Feynman diag o2 self-energy} (\eqref{eq:exc-general-1} and \eqref{eq:E-2-general} in the Appendix), replacing  $\Gb$ with $\GH\se$ and $\Wb$ with $\Wrpa$. Since now the difference between $\vb=\vext+\vH\se$ and the total classical potential $\vext+\vH$ is non vanishing, in principle one also has to take into account the second, fourth, and last \secord{} diagrams. 
However, the difference of densities $n-n_{\rm H}\se$ is at least of first order in the interaction, so the difference of Hartree potentials is at least of second order. For this reason, it does not contribute to the \secord{} diagrams of the self-energy or total xc energy, and we find the same expression as for $\Wb = W$, but with $\Gb\to \GH\se$ and $\Wb\to W^{\rm RPA}$. If one would expand to higher order, instead,  new terms would arise.

\subsection{The total Kohn-Sham potential}
\label{sec:pt-ks}
In our quest to include as much as possible of the interaction effects into $\vb$, while avoiding the complication of a three-arguments effective interaction, a natural choice is
$\vb = \vks = \vext + \vH + \vxc$, where 
the total Kohn-Sham potential $\vks$ is the sum of the total classical potential and the local xc potential, all calculated with the exact charge density $n$. In this case, one will choose $\nb=n$ in Equations \eqref{eq:chi-bar} to \eqref{eq:V-bar}.
Now \eqref{eq:V-bar} is important, since 
$\Vb = v_c + f_{\rm xc}$, the bare Coulomb interaction plus the exchange-correlation kernel $f_{\rm xc}\equiv \dfrac{\delta \vxc}{\delta n}$ of TDDFT. Instead, since the density is exact we have $\chib=\chi$ and $\Pb = P$, the exact reducible and irreducible polarizabilities, respectively. This, together with $\Vb$, leads to $\Wb=\tilde W\equiv v_c + (v_c + f_{\rm xc})\chi v_c$, the exact test charge - test electron screened Coulomb interaction \cite{Singhal1976,Hybertsen1987}. The screening is no longer the one measured in classical experiments, but it is the screening appropriate for the fermions in the system. Following subsection \ref{subsec:gf}, we will use $\Gb=\Gks$, the exact Kohn-Sham Green's function obtained from a Dyson equation with the KS potential.
This leads to the following expression for the self-energy:
\begin{align}
    \Sigmaxc\1(1,2) =& i\Gks(1,2)\tilde W(2,1^+)\nonumber\\
    \Sigmaxc\2(1,2) =& -\Gks(1,\bar 3)\Gks(\bar 3,\bar 4)\tilde W(\bar 4,\bar 3^+)\nonumber \\ 
    &\times\Gks(\bar 4,2)\tilde W(2,1^+)\nonumber\\
    -i\Gks(1,\bar 3)& \vxc^{(1)}(\bar 3)\Gks(\bar 3,2)\tilde W(2,1^+)\nonumber\\
     + i^2 \Gks(1,\bar 4) & \tilde W(\bar 5,1^+)\Gks(\bar 4,\bar 5) \Gks(\bar 5,2) \tilde W(2,\bar 4^+)\nonumber\\
     -i \Gks(1,2) & \tilde W(\bar 5,1^+) \fxc^{(1)}(2,\bar 3)\chi_0(\bar 3,\bar 5)\,.
    \label{eq:S-KS}
\end{align}
Note the appearance of the \firstord{} KS xc potential $\vxc\1$ and kernel $\fxc\1$. As we will see later, it is important to use $\vxc\1$ here, which is $\vxc$ calculated to first order in $\tilde W$, and not the full $\vxc$; similarly for $\fxc$. The importance of this kind of consistency was also pointed out for the diagrammatic expansion of $\vxc$ itself in \cite{Tokatly2001}. One may expect that it influences, for example, the degree to which self-interaction errors, which are found at all levels of approximations involving a screened interaction, are mitigated \cite{Aryasetiawan2012,Isago2025,Isago2026}. The functional derivative of the potential differences in \eqref{eq:sigmaxc-2-g} yields the last row of \eqref{eq:S-KS}, where $\delta \vxc/\delta \bb v $ becomes
\begin{equation}
    \Big (\frac{\delta \vxc}{\delta v_{\rm KS}}\Big )\1=f_{\rm xc}\1\chi_{0}\,.
\end{equation}
Here, we have inserted the exact density $n$ in a chain rule and defined the independent particle response function $\chi_{0} $ to be built with the KS Green's function. As above in the case of $\vxc$, the xc kernel $f_{\rm xc}$ has to be evaluated to \firstord{} in $\tilde W$. 

\begin{figure}
    \centering
         \includegraphics[width=0.35\linewidth]{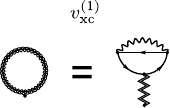}\\
                 \vspace{0.5cm}
    \includegraphics[width=0.7\linewidth]{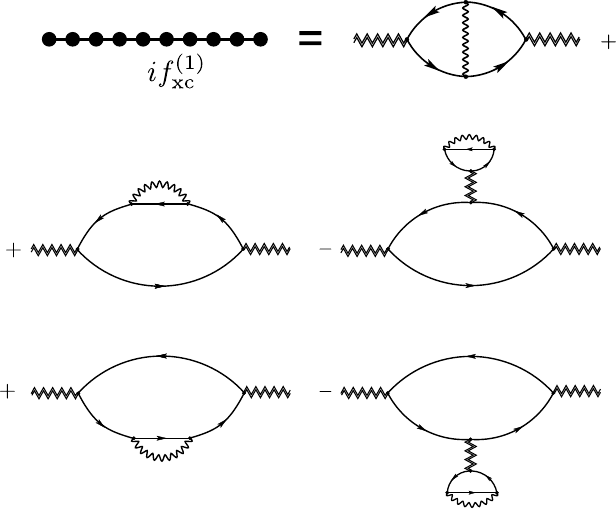}
    \caption{First-order expansion of the xc potential and kernel of TDDFT. A straight arrow indicates $\Gks$, the wiggly line is $i\tilde W$. The double zig-zag line is $i$ times the inverse of the KS independent-particle polarizability, $i\chi_0^{-1}$. The circle stands for $\vxc\1$, given in the first line, and the line with filled circles is $i\fxc\1$ with the electron-hole contribution $i\fxc^{\rm mb}$ (first diagram) and the gap correction contribution $i\fxc^g$ in the next two lines. }
    \label{Feynman diag vxc-fxc o1}
\end{figure}

The \firstord{} KS xc potential is the Optimized Effective Potential-like expression \cite{Sharp1953,Talman1976} with interaction $\tilde W$. This expression can be found, e.g., in Ref.~\cite{Tokatly2001} for the bare interaction and for its extension to RPA screening, and it can be derived for any self-energy (see, e.g., \cite{Vacondio2022}). Here, with $\tilde W$, it reads
\begin{equation}
    \vxc\1(1)= i \chi_0^{-1}(1,\bar 3)\chi_0(\bar 3;\bar 4,\bar 5)\Gks(\bar 4,\bar 5)\tilde W(\bar 5,\bar 4^+)\,,
    \label{eq:vxc1}
    \end{equation}
where $\chi_0(3;4,5)\equiv -i\Gks(3,4)\Gks(5,3^+)$. The diagram is shown in the top row of Fig.~\ref{Feynman diag vxc-fxc o1}.

The \firstord{} xc kernel consists of two parts, $\fxc = \fxc^g+\fxc^{\rm mb}$. The so-called many-body kernel $\fxc^{\rm mb}$ can be directly derived from the Bethe-Salpeter equation \cite{SOT203, MAR203, TOK03, Bruneval2005, GAT07,vonBarth2005}; to first order it reads
\begin{multline}
       \fxc^{{\rm mb}(1)}(1,2) = \chi_0^{-1}(1,\bar 4) \Gks(\bar4,\bar3) \Gks(\bar5,\bar4^{++}) \tilde{W}(\bar3^+,\bar 5)  \\ \times \Gks(\bar3,\bar6) \Gks(\bar6,\bar5) \chi_0^{-1}(\bar 6,2)\,.
       \label{fxc mb o1}
\end{multline}
The first part $\fxc^g$, instead, makes the link between the KS and the Green's functions framework through the difference of the KS independent particle response function and its counterpart $\chi_0^{\rm mb}$ built from the exact dressed $G$. It reads
\begin{multline}
   \fxc^g(1,2) \equiv \chi_{0}^{-1}(1,2) - \chi_{0}^{-1,{\rm mb}}(1,2) \\
   =\chi_{0}^{-1}(1,\bar 3)\bigg( \chi^{\rm mb}_0(\bar 3,\bar 4)  - \chi_0(\bar 3,\bar 4) \bigg) \chi_0^{-1,{\rm mb}}(\bar 4,2)\,. 
\end{multline}
To first order, this yields

\begin{widetext}
\begin{multline}
     \fxc^{g(1)}(1,2)=  \chi_{0}^{-1}(1,\bar 3)
     \Bigg( \Gks(\bar 3, \bar 4) \Gks(\bar 4,\bar 5)  \Gks(\bar 5,\bar 6) \tilde{W}(\bar 6,\bar 5^+)  \Gks(\bar6,\bar3^{++}) 
     +i \Gks(\bar 3, \bar 4) \Gks(\bar 4,\bar 5) \vxc^{(1)}(\bar 5)  \Gks(\bar 5,\bar 3^{++}) \\
     +\Gks(\bar 4, \bar 3^{++}) \Gks(\bar 3,\bar 5)  \Gks(\bar 5,\bar 6) \tilde{W}(\bar 6,\bar 5^+)  \Gks(\bar 6,\bar4) 
     +i \Gks(\bar 4, \bar 3^{++}) \Gks(\bar 3,\bar 5) \vxc^{(1)}(\bar 5)  \Gks(5,4)
     \bigg) \chi_0^{-1}(\bar 4,2)\,.
     \label{fxc gaps o1}
\end{multline}
\end{widetext}
The diagrams for $\fxc^{\rm mb}$ and $\fxc^g$ are given in the first line and in the three subsequent lines of Fig.~\ref{Feynman diag vxc-fxc o1}, respectively. Inserting these terms into \eqref{eq:S-KS} yields the self-energy. 

\begin{center}
\begin{figure*}
    \centering
    \includegraphics[width=\linewidth]{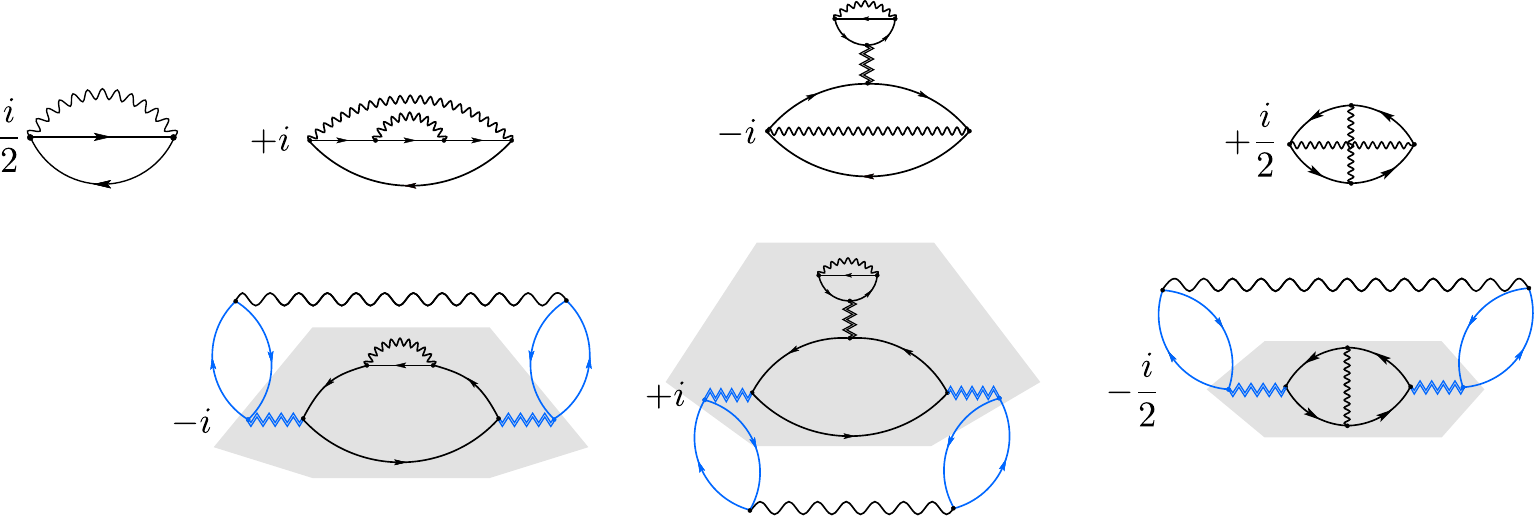}
    \caption{Second-order expansion of the xc energy in terms of KS ingredients. Definitions are the same as in Fig.~\ref{Feynman diag vxc-fxc o1}. The $\vxc\1$ and $\fxc\1$ displayed in Fig.~\ref{Feynman diag vxc-fxc o1} and inserted into Fig.~\ref{Feynman diag o2 self-energy} lead to the \secord{} diagrams for $\Exc$ given here. The first diagram is the \firstord{},  \eqref{eq:sigmaxc-1bb} with $\Gb\to \Gks$ and $\Wb\to \tilde W$. Next come two self-consistency diagrams, and the ordinary \secord{} diagram. The second line contains diagrams stemming from $\fxc^{(1)}$, indicated by the 
grey shaded areas. Bubbles and double zig-zag lines in blue highlight the cancellation between $i\chi_0^{-1}$ contained in $\fxc\1$ and a bubble built with $\Gks$ and contained in the rest of the respective contribution. Finally, each \secord{} diagram in the first line is canceled by its next neighbour in the second line.}
    \label{Feynman diag exc o2}
\end{figure*}
\end{center}

To calculate $\Exc$, we again close the self-energy with a Green's function, including the self-consistency contributions. Equations are given in the Appendix, \eqref{eq:exc-ks-1} to \eqref{eq:exc-ks-final}, and the diagrams are shown in Fig.~\ref{Feynman diag exc o2} (see also, e.g., Ref. \cite{Hellgren2018b} for an expansion in the bare interaction). Here, the fact that certain diagrams appear twice in the general expression Fig.~\ref{Feynman diag o2 self-energy} has already been used to reduce the number of terms. The closing of the loop when going from the self-energy to $\Exc$ creates an additional bubble $-i\chi_0$, such that finally all inverse $i\chi_0^{-1}$ are eliminated. As Fig.~\ref{Feynman diag exc o2} shows, in this way each \secord{} diagram has an identical counterpart of opposite sign, such that the final result is the \firstord{} alone, which is exact
 \cite{elsahili2023total}. 

This cancelation of higher order terms nicely corrobates the consistency requirement put forward in \cite{elsahili2023total}, where the \firstord{} alone was shown to yield exact results under certain conditions. This is very important, not only for immediate applications. Indeed, as we have seen above, the design of a perturbation series involves several choices, and the exact KS starting point yields a rigorous guideline. 

\onecolumngrid

\twocolumngrid
This development is made using exact KS ingredients. In principle, one could also develop in terms of an approximate KS potential, e.g., $\vb=v_{\rm KS}^{\rm LDA}$. In this case the density would not be the exact one, but as in the case of the self-consistent Hartree potential, this would not lead to corrections in the first and \secord{}, because even an approximate $\vxc$ is at least of first order in the interaction. The difficulty, instead, is the fact that for a consistent development, we need $\vxc\1$ and $\fxc\1$ to \firstord{} in $\tilde W$. 
For the exact TDDFT ingredients this is well defined \cite{Tokatly2001,TOK03}. Most approximate functionals, instead, are not diagrammatic, and there is no way to define a \firstord{} unambigously. In the following, we will only consider the case where the TDDFT part is exact, and where the \firstord{} expressions are well defined. In practice, one will then make further approximations at the end, on the resulting expressions: for example, one can use the fact that in the KS-based expansion of the exact $\Exc$ only the \firstord{} contributes, and therefore, approximate directly the \firstord{} expression using $\Gks^{\rm LDA}$ and $W^{\rm ALDA}$, the screened Coulomb interaction calculated in TDDFT using the adiabatic local density approximation. Using exact TDDFT ingredients allows us in the present work to highlight errors that stem from the mere fact that the KS potential is used in place of the self-energy in some finite order expression, independently of approximations to TDDFT.

\section{Illustration: the symmetric Hubbard dimer}
\label{sec:results}

In order to see how these equations perform over a wide parameter range, we use a simple, exactly solvable, model, a Hubbard dimer. This model is widely used to benchmark theoretical developments, including combinations of many-body perturbation theory and other approaches, such as density functional ones (see, e.g., \cite{Kutepov2015,Carrascal2015,Crisostomo2024}). Here, we focus on the symmetric Hubbard dimer \cite{Romaniello2009,Romaniello2012,Carrascal2015,Aryasetiawan2002,Coveney2023}, since its density is exact in all cases by symmetry, which allows us to concentrate on aspects that are not linked to the description of the density. Moreover, correlation effects are enhanced in the symmetric case. The hamiltonian reads \cite{Hubbard1963,Hubbard1964}

\begin{equation}
\hat{H} =  \sum_{i,\sigma}\epsilon_0 \hat{n}_{i\sigma} - t \sum_{<i,j>,i\ne j,\sigma} \hat{c} ^{\dagger}_{i\sigma} \hat{c}_{j\sigma} + U\sum_{i} \hat{n}_{i\uparrow} \hat{n}_{i\downarrow}.
\label{Hubbard hamiltonian}
\end{equation}
where $i,j$ denote the sites $1,2$, the spin $\sigma=\uparrow,\downarrow$, the external on-site potential is
$\epsilon_0$, and 
$t$ is the hopping that is linked to the kinetic energy. 
$U$ is the onsite Coulomb repulsion, and
$\hat{n}_{i\sigma}=\hat{c}^{\dagger}_{i\sigma}\hat{c}_{i\sigma}$ is the particle number operator, where $\hat c$ and $\hat c^{\dagger}$ annihilate and create a fermion, respectively. Using this model allows us to explore the full range of correlation, which can be quantiged by the ratio $U/t$. We will use it at half filling, i.e., with two electrons in a singlet state, which yields non-trivial electron removal and addition features, and we will set $U=4$ eV throughout the illustrations. 

\subsection{Total xc energy}

We start by examining the total xc energy $\Exc$. First-and \secord{} expressions for the Hubbard dimer are given in the Appendices. The results are shown in Fig.~\ref{xc energy plot}. 
\begin{figure*}
    \centering
    \includegraphics[width=0.9\linewidth]{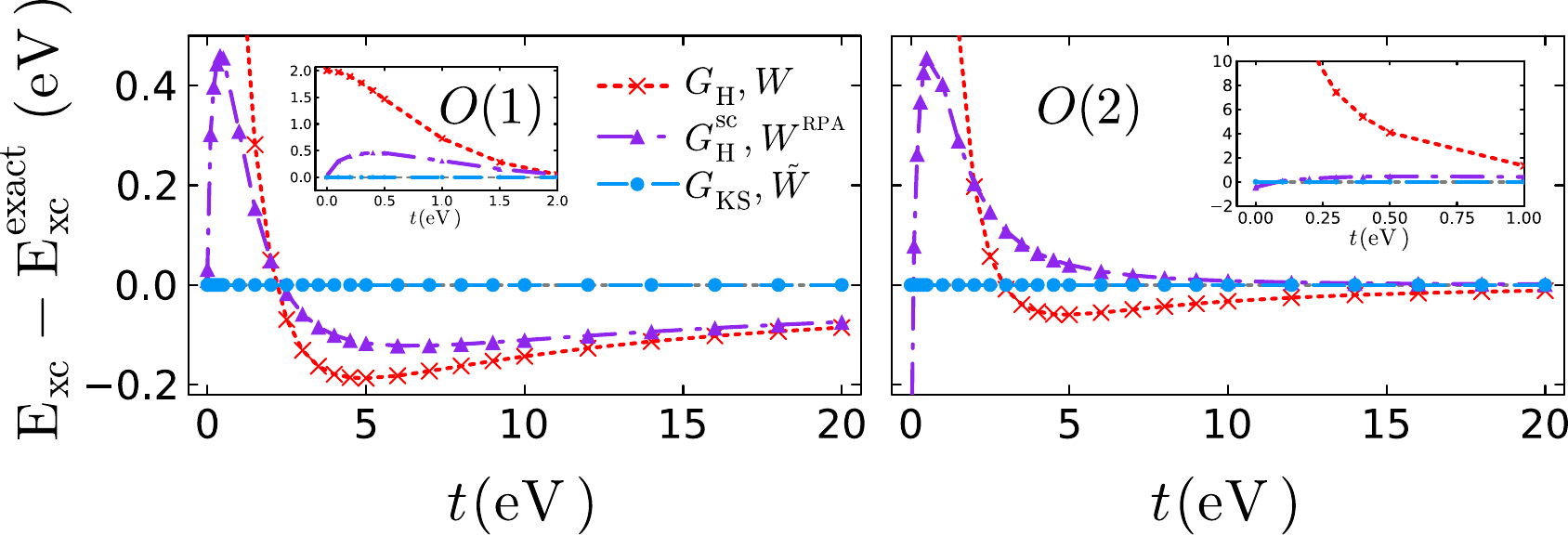}
    \caption{Error in the exchange-correlation contribution to the total energy as a function of $t$ for $U=4 \, \text{eV}$, for different expansions up to first (left) and second (right) orders. Red crosses and violet triangles are results of the expansion in the exact screened Coulomb interaction and the Hartree Green's function with exact density following \eqref{eq:E-2-H} (diagrams given in Fig.~\ref{Feynman diag o2 self-energy}, excluding the second, fourth, and last \secord{} diagrams), and of the expansion in the RPA screened Coulomb interaction and in the Hartree Green's function with self-consistent density (same equations with corresponding replacement of $\Gb$ and $\Wb$), respectively. The results of an expansion in the KS Green's function and the exact test charge-test electron $\tilde W$ (light blue dots) is shown analytically to be exact. The insets are a zoom on a small-$t$ range. }
    \label{xc energy plot}
\end{figure*}

As explained above, the KS based expansion yields the exact result already at first order, and it remains exact also at second order. The two Hartree-based expansions behave similarly at large $t$ in the first-order, but the RPA version is slightly better, and it clearly wins over the exact $W$ at smaller $t$ in first order, and for almost all $t$ when the \secord{} is included. Both expansions improve for larger $t$ when the \secord{} is added to the first one, but the perturbation expansion has problems in a range of small $t$, as one can also see in the zoom in the inset. The small-$t$ results are particularly bad when the exact $W$ is used, where $\Exc$ diverges for $t\to 0$. The fact that standard perturbation theory has convergence problems in a range of strong correlation is of course not astonishing, but it is interesting to observe the very different behavior of the three choices.

\subsection{Expansion of the Kohn-Sham potential}
When one expands around the KS solution up to second order, the knowledge of the exact $\vxc$ as well as its first and \secord{} approximations are required. However, whereas in $\Exc$ all inverse response functions are canceled by bubble diagrams, this is not the case for the calculation of $\vxc$ alone. Therefore, the topmost diagram in Fig.~\ref{Feynman diag vxc-fxc o1}, which is $\vxc^{(1)}$ given in
\eqref{eq:vxc1}, suffers from the fact that $\chi_0$ is not invertible \cite{Niquet2003}. This can be related to the problem that $\vxc$, in order to yield the correct density distribution for fixed particle number, is only determined up to a constant. In the symmetric Hubbard dimer the particle number is the only variable, which means that $\vxc$ is merely a number that sets an energy scale for the self-energy, and that must be determined, but not from the density. We therefore need an additional condition to fix $\vxc$: we will require that the energy of the highest occupied state (HOMO, which is the bonding state denoted by a subscript $b$) be the same, to all orders, whether calculated with the self-energy or with $\vxc$.  

For the exact KS potential, the prescription reads then
    \begin{equation}
        \vxc = {\rm Re}\, (   \Sigma_{\text{xc},b}(\epsilon_{b}) ) \,,
        \label{eq:vxc-exact}
    \end{equation}
with $\epsilon_{b}$ solution of the quasi-particle equation
\begin{equation}
    \epsilon_{b}  = \varepsilon_{\text{H},b} +  {\rm Re}\,( \Sigma_{\mathrm{xc},b}(\epsilon_{b}) )\,.
\end{equation}
Consequently, the order-by order approximation becomes
    \begin{equation}
       \vxc^{(sn)} = {\rm Re}\, (   \Sigma_{\text{xc},b}^{(sn)}(\epsilon^{\text{H}\,(Dn) }_{b}) ) \,,
    \end{equation}
    where $\epsilon^{\text{H}\, (Dn)}_{b}$ is solution of 
 \begin{equation}
    \epsilon^{\text{H}\,(Dn)}_{b}  = \varepsilon_{\text{H},b} +  {\rm Re}\,( \Sigma_{\text{xc},b}^{(sn)}(\epsilon^{\text{H}\,(Dn)}_{b}) )\,.
    \label{eq:qp-H}
\end{equation} 
Note that we distinguish this approximate quasi-particle energy, with superscript $^{\rm H}$, from the solution of
\begin{equation}
    \epsilon^{(Dn)}_{\ell} = \varepsilon_{\text{KS},{\ell}} + {\rm Re}\,( \Sigma_{\text{xc},{\ell}}^{(sn)}(\epsilon^{(Dn)}_{\ell}) ) - v_{\text{xc}}^{(sn)} \,,
    \label{eq:qp-xc}
\end{equation}
that is used for bonding and anti-bonding states ${\ell}$:
whereas to infinite order $\epsilon^{\text{H}\,(Dn)\to\infty}=\epsilon^{(Dn)\to\infty}$,
at any finite order the two differ by higher orders contained in the exact $\epsilon_{\text{KS}}$. We will make use of the improvement brought about by the inclusion of higher orders in $\varepsilon_{\text{KS},{\ell}}$ later on, but for the determination of $v_{\rm xc}^{(sn)}$ one has to start from Hartree, otherwise the problem becomes circular.

More in detail, we determine
    \begin{equation}
       \vxc^{(1)} = {\rm Re}\, (   \Sigma_{\text{xc},b}^{(1)}(\epsilon^{\text{H}\,(D1)}_{b}) ) \,
       \label{eq:vxcs1}
    \end{equation}
    with $    \epsilon^{\text{H}\,(D1)}_{b}  = \varepsilon_{\text{H},b} +  {\rm Re}\,( \Sigma_{\text{xc},b}^{(1)}(\epsilon^{\text{H}\,(D1)}_{b}) )$
and
    \begin{equation}
        \vxc^{(s2)} = {\rm Re}\, \Sigma_{\text{xc},b}^{(1)}(\epsilon_{b}^{\text{H}\, (D2)}) + {\rm Re}\,\Sigma_{\text{xc},b}^{(2)}(\epsilon^{\text{H}\,(D2)}_b) \,
        \label{eq:vxcs2}
    \end{equation}
    with
    $$\epsilon_{b}^{\text{H}\,(D2)}=\varepsilon_{\text{H},b} + {\rm Re}\, \Sigma_{\text{xc},b}^{(1)}(\epsilon_{b}^{\text{H}\,(D2)}) + {\rm Re}\,\Sigma_{\text{xc},b}^{(2)}(\epsilon^{\text{H}\, (D2)}_b).$$
Note that in order to determine the \secord{} the same quasi-particle energy is used in both first-and \secord{} self-energy contributions. This is important in order to guarantee that the expression becomes exact for $n\to \infty$. It also implies for the \secord{} contribution alone that
   $ \vxc^{(2)} = \vxc^{(s2)}-\vxc^{(1)}\neq {\rm Re}\,\Sigma_{\text{xc},v}^{(2)}(\epsilon^{\text{H}\,(D2)})$.

\begin{figure}
    \centering
    \includegraphics[width=0.95\linewidth]{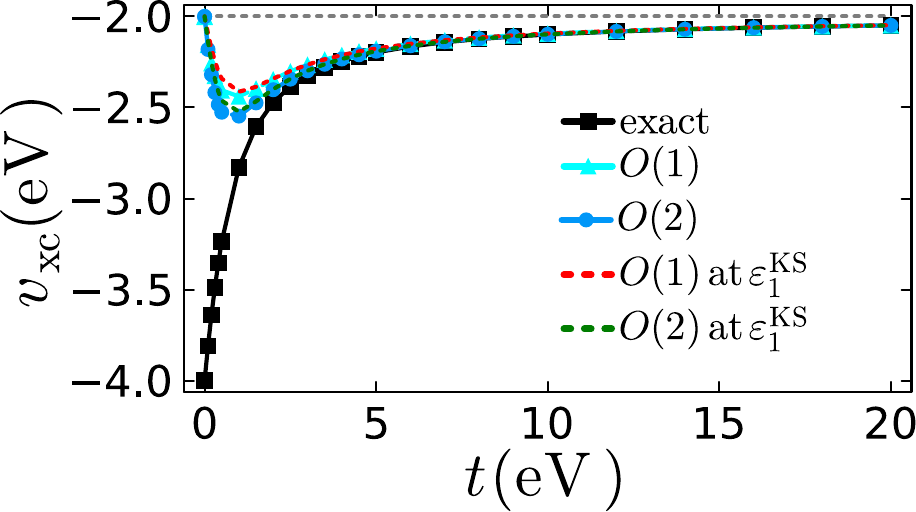}
    \caption{Kohn-Sham exchange-correlation potential $\vxc$ determined from \eqref{eq:vxc-exact} (exact, black curve with squares) and from \eqref{eq:vxcs1} (first-order, light blue with triangles) and \eqref{eq:vxcs2} (\secord{}, darker blue with dots). We also show the results when the self-energy in \eqref{eq:vxcs1} and \eqref{eq:vxcs2} is evaluated at the exact KS energy: this slightly different expansion of $\vxc$ also converges to the exact result, and it may be seen as a valid alternative. The difference is very small, as one can see by comparing the light blue to the red dashed, and the darker blue to the green dashed curve.}
    \label{vxc curves}
\end{figure}

The exact $\vxc$ as well as the first- and \secord{} approximations are shown in Fig.~\ref{vxc curves}. For large $t$, the exact $\vxc$ tends to $-U/2$, which is the exact exchange contribution. For $t\to 0$, it reaches $-U$: each electron is locked on a site, which turns the problem into that of two independent atoms which one electron each, where $\vxc$ exactly cancels the respective Hartree contribution. The first- and \secord{} expansion do well over a large range of $t$, but deviations can be detected for about $t< U$. The \secord{} brings a mild improvement with respect to the first order result, but for $t\to 0$ this improvement vanishes, and both first-and \secord{} tend to the wrong limit $-U/2$. Note that the aim here is not to derive a good approximation for $\vxc$, but to use its low order approximations in order to cancel errors in the corresponding low-order approximations of the self-energy. The good results obtained below for several quantities confirm this strategy. We also checked that with respect to alternative choices our results are stable. In particular, one might also evaluate the self-energy in \eqref{eq:vxcs1} and \eqref{eq:vxcs2} at the exact KS energy: this slightly different expansion of $\vxc$ also converges to the exact result. This leads to an almost indistinguishable $\vxc$, as can be seen in Fig.~\ref{vxc curves}, and we will stick to the first option for all following illustrations. 
    
\subsection{Kinetic energy}
 
\begin{figure*}
    \centering
    \includegraphics[width=0.9\linewidth]{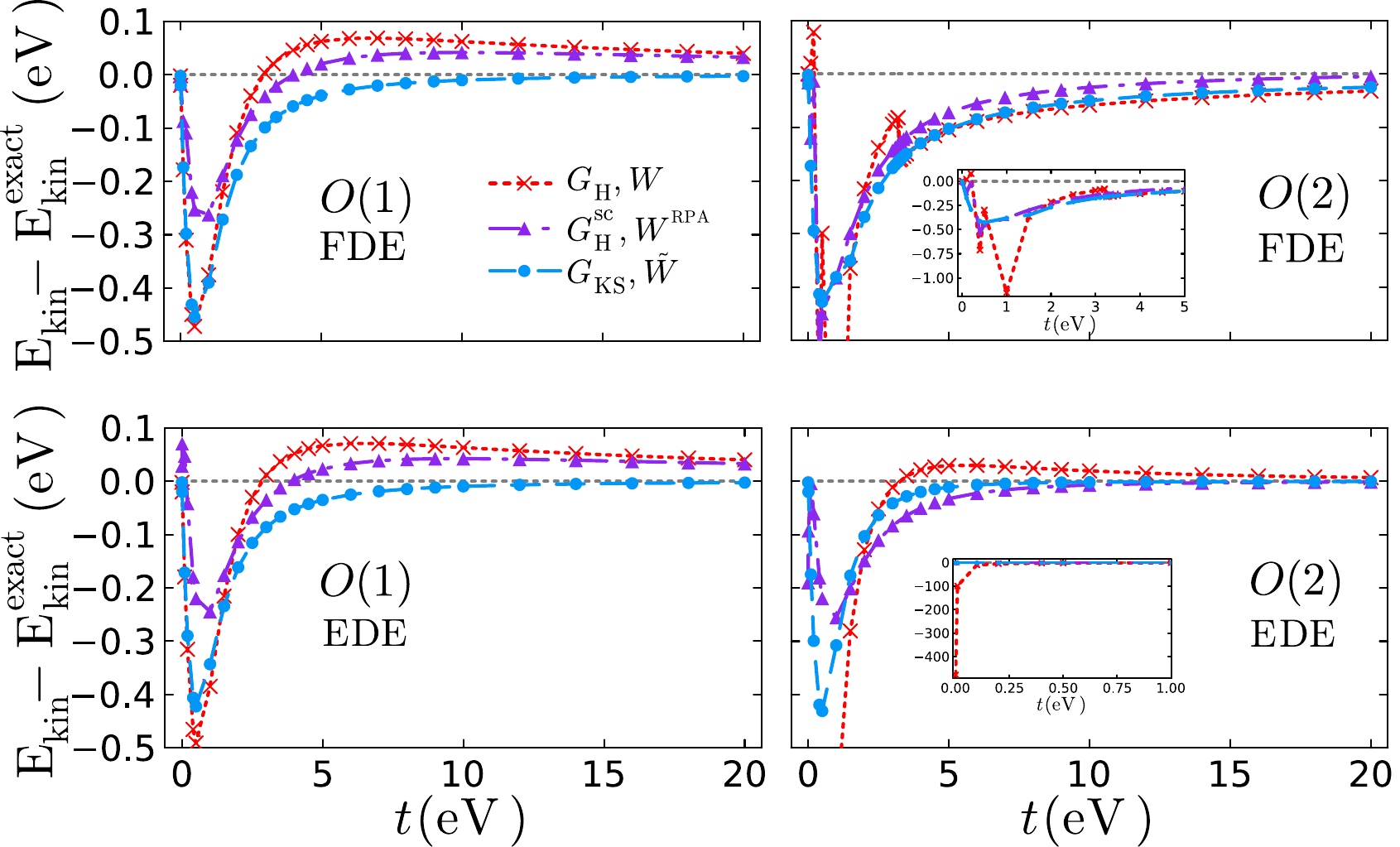}
    \caption{Error in the kinetic energy contribution to the total energy as a function of $t$ for $U=4$ eV, for different expansions up to first (left) and second (right) orders. For the upper panels the Green's function used to calculate the kinetic energy is the solution of the Dyson equation (FDE). For the lower panels, the expansion \eqref{eq:dysonb-ede)} (EDE) was used. Red crosses and violet triangles are results of the expansion of the self-energy in the exact screened Coulomb interaction and the Hartree Green's function with exact density following \eqref{eq:S-H}, and of the expansion in the RPA screened Coulomb interaction and in the Hartree Green's function with self-consistent density (same equations with corresponding replacement of $\Gb$ and $\Wb$), respectively. The results of an expansion in the KS Green's function and the exact test charge-test electron $\tilde W$ (light blue dots) are from the self-energy \eqref{eq:S-KS}. Insets show a zoom on a small $t$ (strong correlation) region. }
    \label{kinetic energy plot}
\end{figure*}

As pointed out in Ref.~\cite{elsahili2023total}, while the $\tilde W$-expansion leads to the exact $\Exc$ already at first order, the same is not true for the kinetic energy $\Ek$, where one has to use explicitly the Green's function, and where other aspects of $G$ are important, compared to $\Exc$. Fig.~\ref{kinetic energy plot} shows  the kinetic energy obtained from the various expansions of the self-energy. In the upper panels, results were calculated from the Green's function $G$  that is solution of the Dyson equation (FDE)\eqref{eq:dysonb-full)}\footnote{Note that the result for $G_{KS}$, $\tilde W$ in the upper left panel is the same as that in Fig.~2 of Ref. \cite{elsahili2023total}, where the solution of the Dyson equation was used.}. In the lower panels, also the Dyson equation itself is expanded consistently, following \eqref{eq:dysonb-ede)} (EDE). 

\begin{figure}
    \centering
    \includegraphics[width=\linewidth]{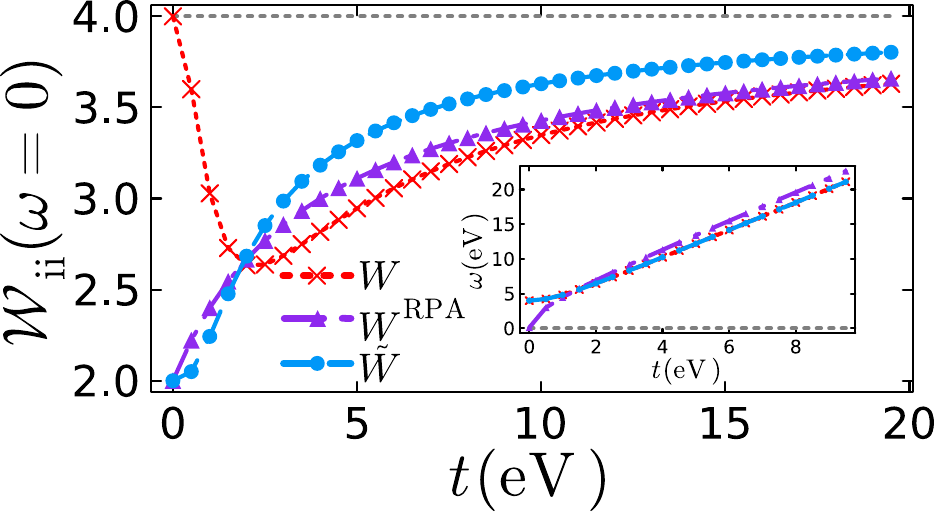}
    \caption{Static values of the diagonal matrix elements in site basis of the screened Coulomb interactions, exact $W$ (red crosses) compared to the RPA (violet triangles) and the test charge-test electron $\tilde W$ (light blue dots). The inset shows the excitation energy in each interaction.}
    \label{fig:screened1}
\end{figure}

The two approximations based on a classical potential show errors in ${\rm E}_{\rm kin}$ that are of similar magnitude and of opposite sign, compared to the respective errors in $\Exc$, both in first and in second order. This leads to error canceling in the total energy. In the $\tilde W$ expansion, the kinetic energy has errors that are not canceled by the errors in $\Exc$, which is exact. Still, concerning the kinetic energy alone and in first order, the $\tilde W$ expansion is the one that approaches the exact result by far the fastest with increasing $t$. The same holds in the \secord{} result, when the consistently expanded Dyson equation (EDE) is used. The \secord{} result obtained from the FDE, instead, is not convincing. Indeed, 
overall the perturbation series seems to converge better when also the Dyson equation is expanded, which supports the suggestion of \cite{Bruneval2019} to update $G$ perturbatively, although the impact is small when one stops at first order.

Concerning the various choices for the perturbation expansion, the RPA and the KS based EDE expansions perform better than the expansion using the exact $W$, similarly to their perfomance for $\Exc$. The KS based expansion wins over the RPA one for $t$ about $t>2$, and it reaches the exact result for $t\to 0$. The RPA-based expansion brings instead some improvement in the remaining range of small $t$.  The relatively bad behavior of the expansion based on the exact $W$ can be traced back to the fact that the exact $W$ is very different from $\Wrpa$ and $\tilde W$; this can be seen in Fig.~\ref{fig:screened1}. The various static screened interactions are similar for large $t$, but the exact $W$ has a minimum around $t =2.2$ eV, and it becomes relatively strong for smaller $t$, which makes the perturbation expansion ill behaved. This confirms the prediction of \cite{Tarantino_2018} that an expansion in the exact screened interaction may suffer from the particular behavior of $W$ in a regime of strong correlation. It is interesting to note, indeed, that the RPA and exact screened test-charge test-charge interactions as a function of correlation strength behave very similarly as in the simple one-point model of Ref.~\cite{Berger_2014,Berger_2014b,Tarantino_2018}. Problems of the spectral function stemming from FDE with an expansion in the exact $W$ will be elucidated more in detail in the next subsection.

\subsection{Spectral properties}

The Green's function and self-energy of the symmetric Hubbard dimer are diagonal in the bonding-antibonding basis. Therefore, the spectral function resulting from a solution of the Dyson equation and using \eqref{spf in terms of the 1-GF} is simply 
\begin{multline}
    A_{\ell}(\omega) = 
     \frac{ \frac{1}{\pi}  {\rm Im}\,\Sigma_{{\rm xc},{\ell}}(\omega)}{(\omega\! -\! \varepsilon_{\ell} \! - \! {\rm Re}\,\Sigma_{{\rm xc},{\ell}}(\omega)\!-\!v_{{\rm cl},{\ell}}\! +\! \vb_{\ell})^2 \!+\! ({\rm Im}\,\Sigma_{{\rm xc},{\ell}}(\omega))^2}\,,
\end{multline}
where $s$ stands for a bonding or antibonding matrix element, and $\varepsilon_{\ell}$ is an eigenvalue belonging to $\vb$. Closest to the Fermi level, when the parenthesis with the real part of $\Sigmaxc(\omega)$ is zero, one has a quasi-particle peak. Satellites, instead, are usually rather associated to peaks in ${\rm Im}\,\Sigma_{{\rm xc},{\ell}}(\omega)$. However, also in a discrete system the position of the satellites can be quite distinct from the peak in the imaginary part of the self-energy:  the satellites in a system without broadening are found as additional solutions of $\omega - \varepsilon_{\ell}-{\rm Re}\,\Sigma_{{\rm xc},{\ell}}(\omega)-v_{{\rm cl},{\ell}}+\vb_{\ell}=0$. When correlation effects are strong, the width of the dispersion of the real part of the self-energy at the position of the peak in the imaginary part is significant. This moves the solution further away from the Fermi level with respect to the peak in ${\rm Im}\,\Sigma_{{\rm xc},{\ell}}(\omega)$, with differences that can be on the scale of eV. An example can be found in the Appendix, Fig.~\ref{spectral features exact vs o2}.

\subsubsection{Quasi-particle energies}

\begin{figure*}
    \centering
    \includegraphics[width=0.9\linewidth]{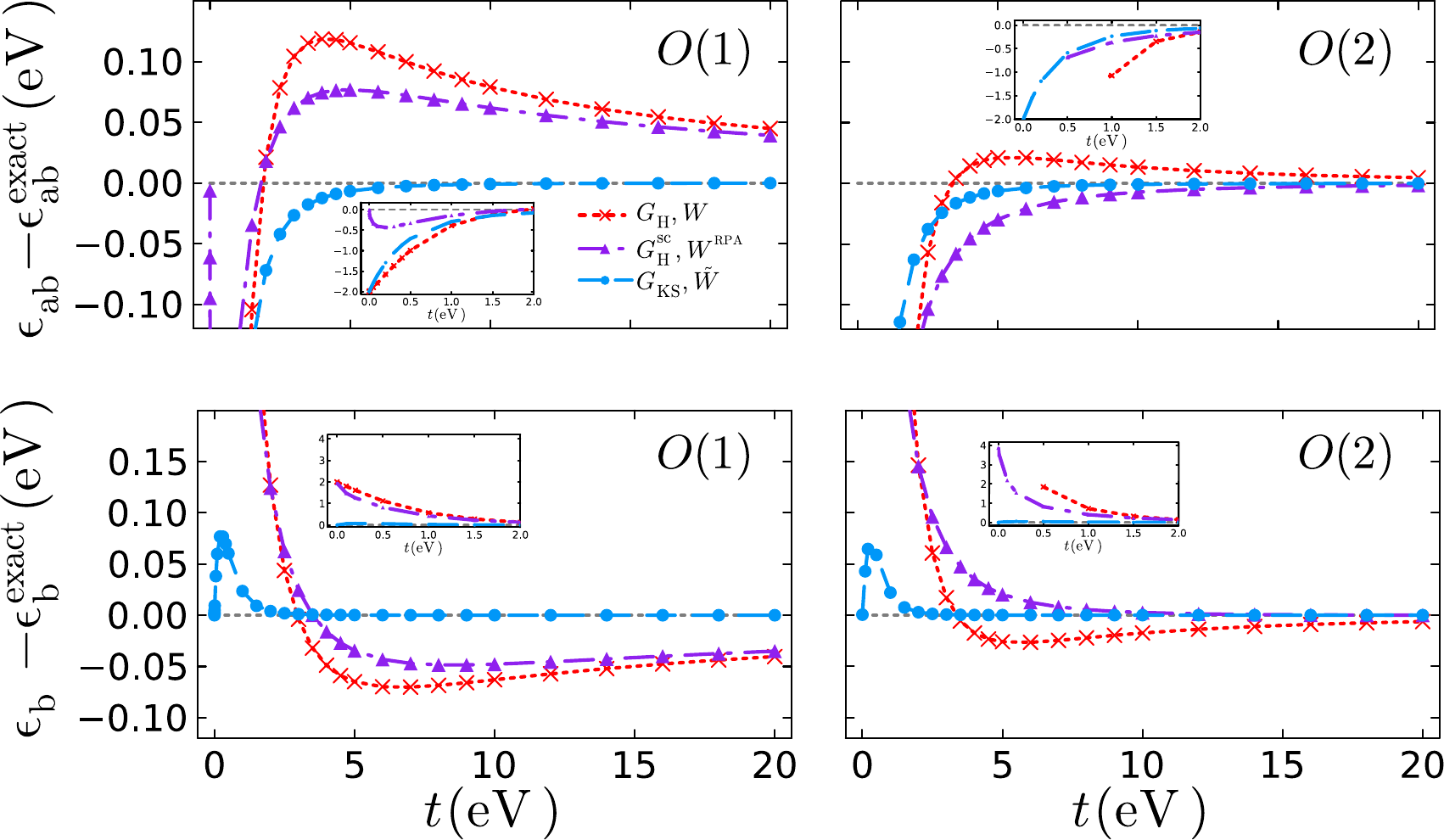}
    \caption{Error in the quasi-particle energies of LUMO (antibonding (ab), top) and HOMO (bonding (b), bottom)
as a function of $t$ for $U=4 \, \text{eV}$, for different expansions up to first (left) and second (right) orders. The Green's function used for this calculation is the solution of the Dyson equation.  Red crosses and violet triangles are results of the expansion of the self-energy in the exact screened Coulomb interaction and the Hartree Green's function with exact density following \eqref{eq:S-H}, and of the expansion in the RPA screened Coulomb interaction and in the Hartree Green's function with self-consistent density (same equations with corresponding replacement of $\Gb$ and $\Wb$), respectively. The results of an expansion in the KS Green's function and the exact test charge-test electron $\tilde W$ (light blue dots) are from the self-energy \eqref{eq:S-KS}. Insets show a zoom on a small $t$ (strong correlation) region.
Missing values indicate a breakdown of the quasi-particle equation.  }
    \label{qp energies plot}
\end{figure*}

Most $GW$ calculations aim at quasi-particle energies. Our results for the Hubbard dimer are presented in Fig.~\ref{qp energies plot}. 
They are obtained from the solution of the Dyson equation, since an expanded Dyson equation is not able to induce quasi-particle shifts. 
For $U/t$ smaller than  $U/t\approx 2$ all errors are smaller than $0.2$ eV. The best results are obtained with the KS-based expansion, for which the HOMO is almost perfect everywhere, and for which also the LUMO is excellent for a large range of larger $t$. The Hartree-based expansions perform less well. In particular, the use of the exact $W$ leads to an error that disappears relatively slowly with increasing $t$. It should, however, be noted that for all approximations in the large $t$ regime the \secord{} result is significantly better than the first-order one.   For $t$ below $2$ eV, except for the HOMO in the KS based expansion, significant errors are found, up to 4 eV in the \secord{} expansion for the LUMO, when $\Wrpa$ is used.
In case of the exact $W$, the spectra in $O(2)$ become so problematic that it is not possible to identify a quasi-particle for small $t$, therefore our curves stop at $t=1$ eV for the LUMO and at $t=0.5$ eV for the HOMO (see inset). The LUMO also has problems when $W^{\rm RPA}$ is used, but only for $t$ smaller than $0.5$ eV. Below, we will look more in detail at the spectral functions.

The KS based expansion is much better behaved, also in the small $t$ range. It shows a, though modest, improvement when going from first to second order, both for the HOMO and for the LUMO. This highlights the KS choice as a valuable way to obtain well behaved perturbative results. For future developments, it could be interesting to introduce two distinct KS potentials for occupied and empty states, simulating the derivative discontinuity. This would go together with a discontinuity in $\fxc$, and it might lead to further improvement for the LUMO. A study in this direction has been performed in Ref. \cite{Hellgren2018}, showing that such a discontinuity would indeed affect the electron affinity. 

\subsubsection{Spectral functions}
\begin{figure*}
    \centering
    \includegraphics[width=0.8\linewidth]{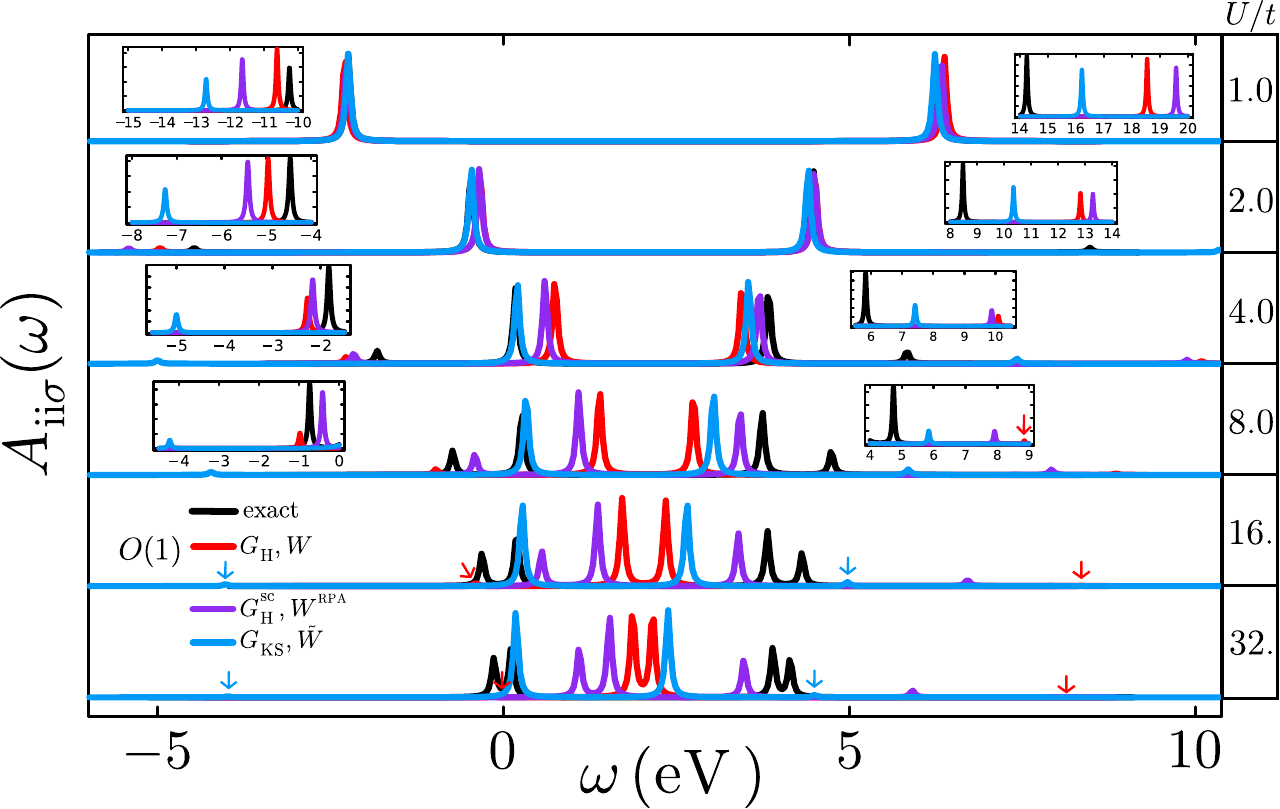}
        \includegraphics[width=0.8\linewidth]{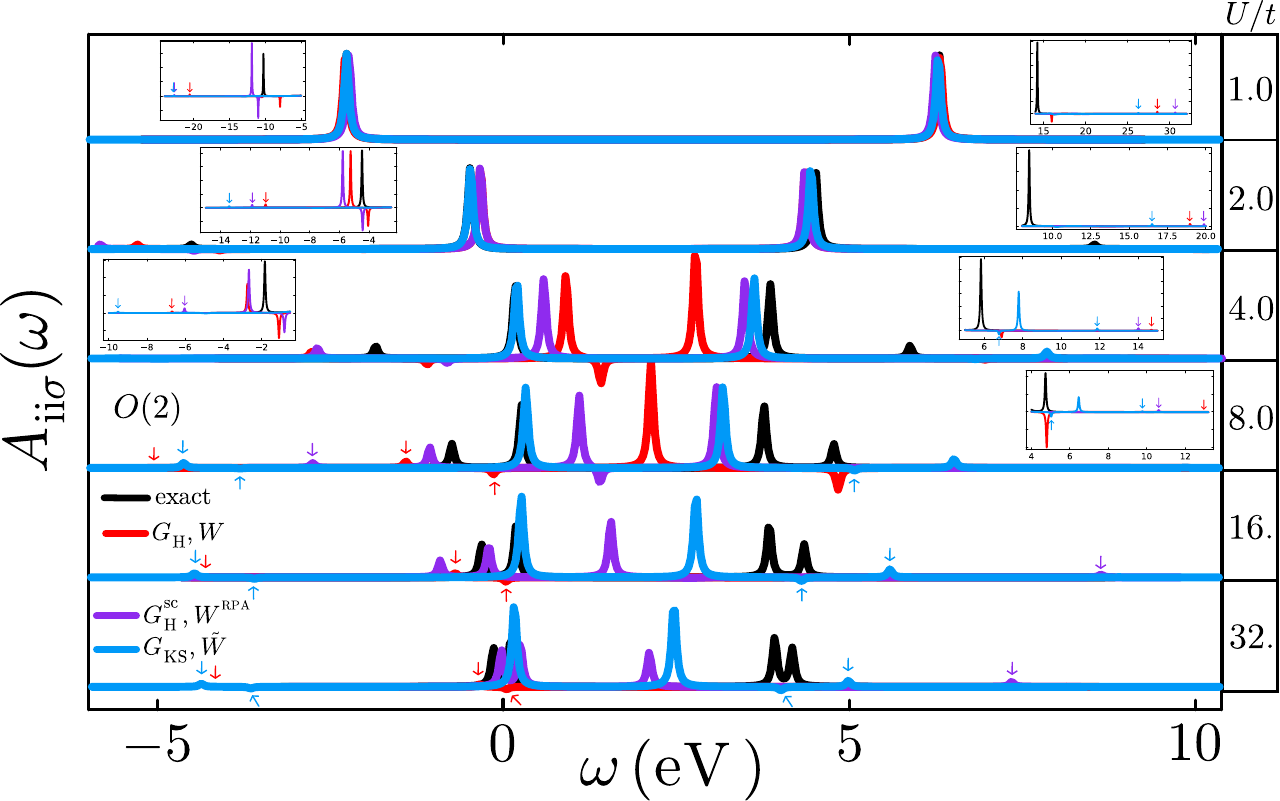}
    \caption{Diagonal matrix element in site and spin basis of the spectral function as a function of the frequency $\omega$. Exact results are shown in black solid lines. Expansions in ($\GH$, $W$), ($\GH^{\text{sc}},\Wrpa$), and $(\Gks, \tilde{W})$ are shown in solid red, violet, and light blue lines, respectively. Different regime of correlation, starting from $U/t = 1.0$ up to $U/t = 32.0$, are given from top to bottom. Insets represent a zoom on the satellites and colored arrows indicate the satellite positions, when they are too weak to be seen on the scale of the figure. Upper panel: expansion up to first order. Lower panel: expansion up to second order. In order to highlight wrong analytic behavior, the results are proportional to the imaginary part of the retarded Green's function, which should be positive everywhere.}
    \label{o1 spectral function}
\end{figure*}

Spectral functions for various values of $U/t$ are shown in Fig.~\ref{o1 spectral function}. The upper panel contains results from an expansion up to order one, and the lower panel has been obtained with an expansion up to second order. 

A few features emerge quite clearly. While the exact QP gap tends to the atomic limit $U=4$ eV, the expansion in the exact $W$ has a tendency to close this gap, at least in the first-order result. In \secord{} no gap can be detected already at $U/t=8$. For even larger $U/t$, no QP peak can be detected at all. This explains why some QP energies are missing in Fig.~\ref{qp energies plot}. Indeed, the \secord{} results becomes erratic, in particular, negative spectral functions are found. Small negative peaks can already be detected at $U/t=1$. The fact that this can happen has been discussed in a general framework \cite{Pavlyukh2016} and is therefore not astonishing. The interesting point is that in the expansion using the exact $W$, the spectral function starts to have severe problems already at $U/t=4$. This can be related to the wrong analytic behavior of ${\rm Im}\,\Sigmaxc$, as explained in \ref{qp calc details} and as illustrated in Fig.~\ref{spectral features exact vs o2}.

The expansion in $\Wrpa$ is better behaved, since significant negative spectral functions and the absence of one of the QP peaks occurs only for much larger values of $U/t$. The expansion in $\tilde W$ yields the most reasonable spectral functions; they are almost exclusively positive, and both the HOMO and LUMO QPs survive even for for very small values of $t$. 

The satellite structures are the most direct fingerprint of correlation effects, since they cannot be obtained in any single particle picture. These are additional structures further away from the Fermi level, whose intensity increases with increasing $U/t$. Below $U/t=4$, the satellites are hardly visible, except for the zoom insets. These insets show that for all $U/t$ the best addition satellite position is obtained in the KS based expansion. Still, the intensity is severely underestimated, and because of the error in the QP position, the distance between QP and satellite, which expresses an excitation energy, is overestimated.  
Instead, the $\tilde W$ expansion gives a very bad description of the removal satellite, whereas the HOMO position is captured very well. This means that again, the energy of the involved neutral excitation is severely overestimated. The $W$ and $\Wrpa$ expansions seem to do better for the removal satellite, but again, relative to the QP the result is unsatisfactory. Note that even with the exact $W$ and $\tilde W$, which both contain the exact excitation energies, the QP-satellite distance may be wrong: 
as outlined above and as can be seen in in Fig.~\ref{spectral features exact vs o2} in the Appendix, the satellites are not exactly at the position of the peak of the imaginary part of the self-energy, which reflects the poles of the screened interaction. Moreover, satellites are known to be sensitive to shifts of the energy scale \cite{Gatti2015,Bruneval2014c}. This scale is set by the $\Gb$ of the various expansions. Since in the Hubbard dimer the removal satellite stems from the anti-bonding contribution of $\Gb$, it is reasonable to think that the fact that $\vxc$ is optimal for the HOMO, but not for the LUMO, plays an important role here. Again, this suggests to introduce in future work the derivative discontinuity of $\vxc$.

\section{Conclusions}
\label{sec:conclusions}

In conclusion, we have discussed a general approach to construct an in principle exact perturbation expansion in terms of a screened interaction and a mean-field Green's function.
This results in a generalization of Hedin's equations, with modified polarizabilities, interactions, and vertex function.

More particularly, we have examined the first- and \secord{} expressions obtained with (i) the self-consistent Hartree Green's function and an RPA screened interaction $W^{\rm RPA}$; (ii) the Hartree Green's function from the exact Hartree potential and the exact test-charge test-charge screened interaction $W$; (iii) the Kohn-Sham Green's function and the corresponding test charge-test electron screened interaction $\tilde W$, where ingredients from TDDFT are used. Option (iii), when used consistently, yields the exact exchange-correlation contribution $\Exc$ to the total potential energy already at first order, and the \secord{} vanishes. In order to obtain this result, it is important to determine the TDDFT ingredients at the corresponding order, instead of using the exact ones everywhere. This serves as a guideline for the general case.

We have tested the three expansions for the symmetric Hubbard dimer over a wide parameter range. Option (iii) performs systematically best, which speaks for the fact that this is a promising direction, and which corrobates the good results found in \textit{ab initio} calculations for real materials, though with approximate TDDFT ingredients
\cite{Chen2015,Tal2021,Abdallah2024,Olsen2019}. Option (ii), the use of the exact test-charge test-charge screened interaction, appears to be the most problematic: of the three choices, it yields the worst results for $\Exc$, and no reliable Green's function can be obtained from the Dyson equation even when $t$ is of the order of $U/4$, due to strong features in the imaginary part of the self-energy with wrong analytic behaviour. This, in turn, reflects in the kinetic energy, quasi-particle energies, and overall spectral functions. Interestingly, this confirms the prediction of a very simple model, where space, spin and time are collapsed in one point
and only the mathematical structure of the equations is
retained \cite{Berger_2014,Tarantino_2018}, which implies that the problem is deeply rooted in the structure of the equations. Option (i), which uses the RPA screened interaction, is maybe the most widely used flavour in \textit{ab initio} calculations, though in general used with other input Green's functions, such as the KS one. It clearly beats the expansion (ii) in the exact screened interaction, and it performs very well for the kinetic energy, especially in the range of small $t$, when the expanded Dyson equation is used. Second-order QP energies are much better than first-order ones, except for small $t$, where the perturbation series breaks down. 

None of the approximations yields good satellite spectra. This is in line with long-standing experience concerning the fact that the solution of the Dyson equation favors good QP energies, but not good satellite positions \cite{Blomberg1972,Bergerse1973,Reining2018}; the satellites suffer from the fact that the solution of the Dyson equation with a low-order self-energy is not order-by-order consistent, and one should rather treat the QP part of the spectrum and and satellites separately. This also nicely illustrates the fact that even very wrong spectral functions can lead to good kinetic and xc energies: these quantities relie on very different aspects of the Green's function, and there is no reason to think that one and the same approximation will describe all of them in an optimal way. 

Besides self-consistency diagrams that are included in a consistent expansion, we have not touched upon the question of self-consistency here in order to keep the picture manageable. In practical calculations, where the exact KS potential is not known, self-consistency may play different roles, for example, it may correct the approximate KS HOMO, or it may create a density that does not suffer from self-interaction errors induced by the approximate KS potential. Approaches such as QS$GW$ \cite{Faleev2004,Kotani2007} play therefore an important role in making $GW$ calculations successful in practice, and an additional effort is required to merge the insight from general developments, model results and \textit{ab initio} calculations. Some simple actions can readily be carried out, such as the alignment of the HOMO as already proposed by Hedin \cite{Hedin1999}, which would implicitly correct the starting KS potential. 

\section*{Data/Software Availability}
Calculations for this article have been performed using the code \href{https://gitlab.com/aelsahili-group/mbpt-in-hubbard-dimer}{MBPT in Hubbard dimer} written in Julia, and developed by the authors.

All self-energy integrals and exchange-correlation energies were evaluated analytically. The Green's functions were then obtained numerically by solving the Dyson equation, both in its full and expanded forms. These Green's functions were subsequently used to compute the kinetic energy by numerically evaluating the integral in Eq.~\eqref{ek integral dimer} using the \texttt{quadgk} library, which is based on Gauss--Kronrod quadrature schemes. The spectral functions were also computed from the resulting Green's functions.
The quasi-particle energies were obtained by numerically solving the quasi-particle equations using the \texttt{Roots} library.

All figures presented in this work were generated using the \texttt{Plots} package with the \texttt{GR} backend.

All results for this article can be produced, on the fly, with the \texttt{publication\_release} version of the code \cite{code}.

\section*{Acknowledgments}
The authors acknowledge the fruitful discussions with Matteo Gatti. This project has received funding from the European
Union’s Horizon 2020 research and innovation programme
under grant agreement no. 800945 -- NUMERICS -- H2020-
MSCA-COFUND-2017.
\section{APPENDIX}

\subsection{Validity of chain rules}
\label{sec:app-chain}
A chain rule, such as the insertion of $\vb$ into $\dfrac{\delta G^{-1}}{\delta \vext}$ leading to $\dfrac{\delta G^{-1}}{\delta \vb}\dfrac{\delta \vb}{\delta \vext}$ in \eqref{eq:sigmaxc-vbar}, can be used safely when $G^{-1}[\vext]$ can be understood as a functional $\tilde G^{-1}[\vb]$, and when $\vb$ is functional of $\vext$ \cite{Strinati1988}. If $\vext$ is functional of $\vb$ then  
$\tilde G^{-1}[\vb]=G^{-1}[\vext[\vb]]$ fulfills the first requirement. In other words, when there is a one-to-one relation between $\vext$ and $\vb$, everything is safe.  

$\vext$ determines all the potentials $\vb$ that we are using, because it uniquely determines both the exact density and the self-consistent Hartree density. In turn, these ingredients fully determine our $\vb$.

It is more difficult to show in general that a given $\nu$ uniquely determines the external potential, except for the case $\nu=\vks$, where the KS potential determines the exact density, for which the Hohenberg-Kohn theorem establishes the one-to-one relation with $\vext$. The case of the self-consistent Hartree potential is analogous. 

In the case of the total classical potential with \textit{exact} density, the density and the potential are not related in this simple self-consistent way, and the one-to-one relation may not hold. This subtle question will be deepened elsewhere; for the present work, we simply explore the results obtained with the widely used assumption that we may use this chain rule, but keeping in mind that we have not proved it.

Finally, note that the one-to-one relation between $\vext$ and the densities $n$ or $\nb$ used in this work also justifies the chain rule with the density in \eqref{eq:chain-barn}.

\subsection{General equations \label{detailed deriv}}

The EDE, Eq.~\eqref{eq:dysonb-ede)}, reads 
\begin{multline}
G^{(s2)}(1,2)= \Gb(1,2) \\
+ \Gb(1,\bar3) \bigg( \Sigmaxc^{(1)}(\bb3,\bb4)  - \big(\vb(\bb3,\bb4) -\vcl(\bb3,\bb4) \big)^{(1)} \delta(\bb3,\bb4) \bigg) \Gb(\bb4,2) \\ +   \Gb(1,\bar3) \bigg( \Sigmaxc^{(2)}(\bb3,\bb4) - ( \vb(\bb3,\bb4)-\vcl(\bb3,\bb4))^{(2)}\delta(\bb3,\bb4) \bigg) \Gb(\bb4,2)\\
+ \Gb(1,\bar3) \bigg( \Sigmaxc^{(1)}(\bb3,\bb4)  - (\vb(\bb3,\bb4) -\vcl(\bb3,\bb4))^{(1)} \delta(\bb3,\bb4) \bigg) \Gb(\bb4,\bb5)\!\! \\  \times \bigg( \Sigmaxc^{(1)}(\bb5,\bb6)\! -\! (\vb(\bb5,\bb6) \!-\!\vcl(\bb5,\bb6))^{(1)} \delta(\bb5,\bb6) \bigg) \Gb(\bb6,2)\,.
\label{eq:kinetic-expand}
\end{multline}

The total xc energy to second order, corresponding to the diagrams in Fig. \ref{Feynman diag o2 self-energy}, reads

\begin{widetext}
\begin{equation}
   \Exc\1=\frac{1}{2}\Gb(1,\bar 3)\Wb(\bar 3,1^+)\Gb(\bar 3,1^{++})\,,
   \label{eq:exc-general-1}
\end{equation}

    \begin{multline}
    \Exc\2=
    \frac{i}{2}\Gb(1,\bar 4)\Gb(\bar 4,\bar 5)\Wb(\bar 5,\bar 4^+)\Gb(\bar 5,\bar 3)\Wb(\bar 3,1^+)\Gb(\bar 3,1^{++}) 
    -\frac{1}{2} \Gb(1,\bar 4)(\vb\1(\bar 4)-\vcl\1(\bar 4))\Gb(\bar 4,\bar 3) \Wb(\bar 3,1^+)\Gb(\bar 3,1^{++}) \\
    +
    \frac{i}{2}\Gb(1,\bar 3)\Wb(\bar 3,1^+)\Gb(\bar 3,\bar 4)\Gb(\bar 4,\bar 5)\Wb(\bar 5,\bar 4^+)\Gb(\bar 5,1^{++}) 
    - 
    \frac{1}{2}\Gb(1,\bar 3)\Wb(\bar 3,1^+)\Gb(\bar 3,\bar 4)(\vb\1(\bar 4)-\vcl\1(\bar 4))\Gb(\bar 4,1^{++}) \\
    +\frac{i}{2}\Gb(1,\bar 4) \Wb(\bar 5,1^+)\Gb(\bar 4,\bar 5) \Gb(\bar 5,\bar 3) \Wb(\bar 3,\bar 4^+)\Gb(\bar 3,1^{++}) 
     - \frac{1}{2} \Gb(1,\bar 3)\Wb(\bar 5,1^+)\Big [\frac{\delta(\vb(\bar 3)-\vcl(\bar 3))}{\delta \vb(\bar 5)}\Big ]\1 \Gb(\bar 3,1^{++})\,.
        \label{eq:E-2-general}
\end{multline}
\end{widetext}

The expansion using the total classical potential yields

\begin{align}
    &\Sigmaxc\1(1,2) = i\GH(1,2)W(2,1^+)\nonumber\\
    &\Sigmaxc\2(1,2)  = -\GH(1,\bar 3)\GH(\bar 3,\bar 4)W(\bar 4,\bar 3^+)\GH(\bar 4,2) W(2,1^+)\nonumber \\
    &-\GH(1,\bar 4) W(\bar 5,1^+)G(\bar 4,\bar 5) \GH(\bar 5,2) W(2,\bar 4^+)  \,.
    \label{eq:S-H}
\end{align}

And for the xc energy: 
\begin{equation}
   \Exc\1=\frac{1}{2}\GH(1,\bar 3)W(\bar 3,1^+)\GH(\bar 3,1^{++}) \,,
   \label{eq:exc-H-1}
\end{equation}
\begin{multline}
    \Exc\2 = \frac{i}{2}\GH(1,\bar 4)\GH(\bar 4,\bar 5)W(\bar 5,\bar 4^+)\GH(\bar 5,\bar 3) W(\bar 3,1^+)\\ \times  \GH(\bar 3,1^{++}) 
    + \frac{i}{2}\GH(1,\bar 3)W(\bar 3,1^+)\GH(\bar 3,\bar 4)\GH(\bar 4,\bar 5)\\ \times W(\bar 5,\bar 4^+)\GH(\bar 5,1^{++}) 
    +\frac{i}{2}\GH(1,\bar 4) W(\bar 5,1^+)\GH(\bar 4,\bar 5) \\ \times \GH(\bar 5,\bar 3) W(\bar 3,\bar 4^+)\GH(\bar 3,1^{++})\,.
    \label{eq:E-2-H}
\end{multline}

Eqs.~\eqref{eq:S-H}, \eqref{eq:exc-H-1}, and $\eqref{eq:E-2-H}$ are the same for the case of the RPA expansion, where $\Gb = \GH^{\text{sc}}$ and $\Wb = \Wrpa$. For the case of the expansion around KS, we get:

\begin{equation}
 \Exc\1=\frac{1}{2}\Gks(1,\bar 3)\tilde W(\bar 3,1^+)\Gks(\bar 3,1^{++})\,,
 \label{eq:exc-ks-1}
\end{equation}

\begin{multline}
    \Exc\2=
    \frac{i}{2}\Gks(1,\bar 4)\Gks(\bar 4,\bar 5)\tilde W(\bar 5,\bar 4^+)\Gks(\bar 5,\bar 3) \\ \times\tilde W(\bar 3,1^+)\Gks(\bar 3,1^{++}) \\
    -\frac{1}{2} \Gks(1,\bar 4)\vxc\1(\bar 4)\Gks(\bar 4,\bar 3) W(\bar 3,1^+)\Gks(\bar 3,1^{++})  \\
    +\frac{i}{2}\Gks(1,\bar 3)\tilde W(\bar 3,1^+)\Gks(\bar 3,\bar 4)\Gks(\bar 4,\bar 5)\\ \times \tilde W(\bar 5,\bar 4^+) \Gks(\bar 5,1^{++}) \\
    -\frac{1}{2}\Gks(1,\bar 3)\tilde W(\bar 3,1^+)\Gks(\bar 3,\bar 4)\vxc\1(\bar 4)\Gks(\bar 4,1^{++}) \\
    + \frac{i}{2}\Gks(1,\bar 4) \tilde W(\bar 5,1^+)\Gks(\bar 4,\bar 5) \Gks(\bar 5,\bar 3)\\ \times  \tilde W(\bar 3,\bar 4^+)\Gks(\bar 3,1^{++}) \\
     -\frac{1}{2} \Gks(1,\bar 3) \fxc^{(1)}(\bar 3,\bar 4) \chi_0(\bar 4,\bar 5)\tilde{W}(\bar 5,1^+)\Gks(\bar 3,1^{++})\,.
        \label{eq:E-KS}
\end{multline}
 Using the explicit expressions for the \firstord{} xc potential and kernel, the total xc energy becomes
\begin{widetext}
 
\begin{multline}
    \Exc\2= \frac{i}{2}\Gks(1,\bar 4)\Gks(\bar 4,\bar 5)\tilde W(\bar 5,\bar 4^+)\Gks(\bar 5,\bar 3) \tilde W(\bar 3,1^+)\Gks(\bar 3,1^{++}) 
   \, - \, \frac{1}{2} \Gks(1,\bar 4) \vxc\1(\bar 4)\Gks(\bar 4,\bar 3) \tilde W(\bar 3,1^+)\Gks(\bar 3,1^{++})\\
    +\frac{i}{2}\Gks(1,\bar 3)\tilde W(\bar 3,1^+)\Gks(\bar 3,\bar 4)\Gks(\bar 4,\bar 5)\tilde W(\bar 5,\bar 4^+)\Gks(\bar 5,1^{++}) 
    -\frac{1}{2}\Gks(1,\bar 3)\tilde W(\bar 3,1^+)\Gks(\bar 3,\bar 4)
    \vxc\1(\bar 4) \Gks(\bar 4,1^{++}) \\
    +\frac{i}{2}\Gks(1,\bar 4) \tilde W(\bar 5,1^+)\Gks(\bar 4,\bar 5) \Gks(\bar 5,\bar 3) \tilde W(\bar 3,\bar 4^+)\Gks(\bar 3,1^{++}) \\
     - \frac{1}{2} \Gks(1,\bar 3) \chi^{-1}_0(3,\bar 7) \Gks(\bar 7,\bar 6) \Gks(\bar 8,\bar 7^{++}) \tilde{W}(\bar 6^+,\bar 8) \Gks(\bar 6, \bar 9) \Gks(\bar9,\bar8) \chi^{-1}_0(\bar 9,\bar 4)\chi_0(\bar 4,\bar 5) \tilde{W}(\bar 5,1^+) \Gks(\bar 3, 1^{++}) \\
    - \frac{1}{2} \Gks(1,\bar 3) \chi_0^{-1}(\bar3,\bar6) \Gks(\bar 6,\bar 7) \Gks(\bar7,\bar8) \Gks(\bar 8,\bar9) \tilde W(\bar 9,\bar 8^+)  \Gks(\bar 9,\bar6^{++}) \chi_0^{-1}(\bar 7,\bar 4) \chi_0(\bar 4,\bar 5) \tilde W(\bar 5,1^+) \Gks(\bar 3,1^{++}) \\
   - \frac{i}{2} \Gks(1,\bar 3) \chi_0^{-1}(\bar3,\bar6) \Gks(\bar 6,\bar 7) \Gks(\bar7,\bar8) \vxc^{(1)} (8)  \Gks(\bar 8,\bar6^{++}) \chi_0^{-1}(\bar 7,\bar 4) \chi_0(\bar 4,\bar 5) \tilde W(\bar 5,1^+) \Gks(\bar 3,1^{++}) \\    
     - \frac{1}{2} \Gks(1,\bar 3) \chi^{-1}_0(\bar3,\bar6) \Gks(\bar 7,\bar 6^{++}) \Gks(\bar 6,\bar 8) \Gks(\bar8,\bar9) \tilde W(\bar 9,\bar 8^+) 
    \Gks(\bar 9,\bar 7) \chi_0^{-1}(\bar 7,\bar 4) \chi_0(\bar 4,\bar 5) \tilde W(\bar 5,1^+) \Gks(\bar 3,1^{++}) \\
    - \frac{i}{2} \Gks(1,\bar 3) \chi^{-1}_0(\bar3,\bar6) \Gks(\bar 7,\bar 6^{++}) \Gks(\bar 6,\bar 8) \vxc^{(1)}(\bar 8) 
  \Gks(\bar 8,\bar 7) \chi_0^{-1}(\bar 7,\bar 4) \chi_0(\bar 4,\bar 5) \tilde W(\bar 5,1^+) \Gks(\bar 3,1^{++})\,.
        \label{eq:E-KS}
\end{multline}

This simplifies to

\begin{multline}
    \Exc\2=
    \frac{i}{2}\Gks(1,\bar 4)\Gks(\bar 4,\bar 5)\tilde W(\bar 5,\bar 4^+)\Gks(\bar 5,\bar 3) \tilde W(\bar 3,1^+)\Gks(\bar 3,1^{++})
    \, - \, \frac{1}{2} \Gks(1,\bar 4)
    \vxc\1(\bar 4)\Gks(\bar 4,\bar 3) \tilde W(\bar 3,1^+)\Gks(\bar 3,1^{++}) \\
    +\frac{i}{2}\Gks(1,\bar 3)\tilde W(\bar 3,1^+)\Gks(\bar 3,\bar 4)\Gks(\bar 4,\bar 5)\tilde W(\bar 5,\bar 4^+)\Gks(\bar 5,1^{++}) 
    \, - \, \frac{1}{2}\Gks(1,\bar 3)\tilde W(\bar 3,1^+)\Gks(\bar 3,\bar 4)
    \vxc\1(\bar 4) \Gks(\bar 4,1^{++}) \\
    +\frac{i}{2}\Gks(1,\bar 4) \tilde W(\bar 5,1^+)\Gks(\bar 4,\bar 5) \Gks(\bar 5,\bar 3) \tilde W(\bar 3,\bar 4^+)\Gks(\bar 3,1^{++})
     \, -\,  \frac{i}{2} \Gks(1,\bar 4) \tilde{W}(\bar 5, 1^+) \Gks(\bar4, \bar 5)\Gks(\bar 5,\bar 3) \tilde W(\bar 4^+,\bar 3) \Gks(\bar 3,1^{++}) \\
     -\frac{i}{2} \Gks(1,\bar 3) \tilde{W}(\bar3,1^+) \Gks(\bar3,\bar4) \Gks(\bar4,\bar5)\tilde W(\bar5,\bar4^+) \Gks(\bar5,1^{++}) 
     + \frac{1}{2} \Gks(\bar 1,\bar 3) \tilde W(\bar 3,1^+) \Gks(\bar 3,\bar4) \vxc^{(1)}(\bar 4) \Gks(\bar 4,1^{++}) \\
     -\! \frac{i}{2} \Gks(1,\bar 4) \Gks(\bar 4,\bar 5) \tilde W(\bar 5,\bar 4^+) \Gks(\bar 5,\bar 3) \tilde W(\bar3,1^+) \Gks(\bar 3,1^{++})
    \! +\! \frac{1}{2} \Gks(1,\bar 4) \vxc^{(1)}(\bar 4) \Gks(\bar 4,\bar 3) \tilde W(\bar3,1^+) \Gks(\bar 3,1^{++}),
    \label{eq:exc-ks-final}
\end{multline}
\end{widetext}
where we have renamed arguments such that one can see the cancelations clearly and that $\Exc\2 = 0 $ when $\Gb = \Gks$ and $\Wb = \tilde{W}$. 

\subsection{Expressions for the symmetric Hubbard dimer}

The exact solutions for the Hubbard dimer are given in several references, see, e.g., \cite{Romaniello2009, Romaniello2012}. Here, we report all formulas used in this paper to illustrate our findings.

\subsubsection{Green's functions and screened interactions}

The non-interacting Green's functions are $4\times4$ matrices with $2\times2$ equivalent blocks for spin up and down,
\begin{equation}
    \Gb(\omega) =\begin{pmatrix}
        \Gb_{11\uparrow}(\omega) &  \Gb_{12\uparrow}(\omega) & 0 & 0\\
        \Gb_{21\uparrow}(\omega) &  \Gb_{22\uparrow}(\omega) & 0 & 0\\
         0 & 0 & \Gb_{11\downarrow}(\omega) &  \Gb_{12\downarrow}(\omega)\\
          0 & 0 & \Gb_{21\downarrow}(\omega) &  \Gb_{22\downarrow}(\omega)\\
    \end{pmatrix}\,,
\end{equation}

where
\begin{multline}
    \Gb_{ij\sigma}(\omega) = \frac{1}{2} \bigg( \frac{1}{\omega-\varepsilon_b -i\eta} + \frac{1}{\omega-\varepsilon_{ab} +i\eta} \bigg)\,
\end{multline}
with $\Gb_{ij\uparrow}(\omega) = \Gb_{ij\downarrow}(\omega) $.
 $\varepsilon_b = \epsilon_0-t+\bar \Delta $ and $\varepsilon_{ab} = \epsilon_0+t+\bar \Delta $ are the highest occupied and lowest unoccupied molecular orbitals energies, respectively, with $\bar \Delta$ a shift that depends on the choice of $\Gb$. For $\Gb = \GH = \GH^{\text{sc}}$,  $\varepsilon_b = \varepsilon_b^{\text{H}}$,  $\varepsilon_{ab} = \varepsilon_{ab}^{\text{H}}$, where $\bar \Delta = \Delta_{\text{H}} = U$. For $\Gb = \Gks$, $\varepsilon_b = \varepsilon_b^{\text{\tiny KS}}$ and  $\varepsilon_{ab} = \varepsilon_{ab}^{\text{\tiny KS}}$, where $\bar \Delta = \Delta_{\mathrm{KS}} = 2t - \frac{c-U}{2}$ with $c = \sqrt{16t^2 + U^2}$. \\

The screened interactions are
\begin{multline}
\Wb_{ij}(\omega) = U\delta_{ij} + \underbrace{(-1)^{i-j}X}_{\text{only when}\, \Wb = \tilde W} + \bar{A}\bigg( \frac{1}{\omega-\bar{\omega}_p+i\eta} \\- \frac{1}{\omega+\bar{\omega}_p-i\eta} \bigg) \,,
\end{multline}
where for $\Wb = W$, $\bar \omega_p = \omega_p = \frac{c+U}{2} $ and $\bar A = A= \frac{2U^2}{a^2}$ with $a^2 = 2\big( \frac{16t^2}{(c-U)^2} + 1 \big)$. For $\Wb = \Wrpa$, $\bar \omega_p = \omega^{\mathrm{RPA}}_p = \sqrt{4t^2 + 4Ut} $ and $\bar A = A^{\mathrm{RPA} } = \frac{U^2t}{\omega^{\text{\tiny RPA}}_p}$. For $\Wb = \tilde W$,  $\bar \omega = \tilde \omega_p = \frac{c+U}{2}$, $X = - \frac{U}{2} +\frac{2U\tilde \omega_p}{2ta^2} $ and $\bar A  = \tilde A =\frac{U( \tilde \omega_p^2 - 4t^2)}{2ta^2}  $.

\subsubsection{First-order expansions}
The \firstord{} self-energy formula in the $ij$ basis of the dimer is 
\begin{equation}
    \Sigma^{(1)}_{\text{xc} \,ij\sigma}(\omega) = \frac{i}{2\pi} \int d\omega' \,  \Gb_{ij\sigma}(\omega'+\omega) \Wb_{ji}(\omega') e^{i\omega'\eta}\,.
\end{equation}
The solutions are
\begin{multline}
    \Sigma^{(1)}_{\text{xc} \,ij\sigma}(\omega) = -\frac{U}{2} \delta_{ij} \underbrace{-(-1)^{i-j}\frac{X}{2}}_{ \text{only when}\, \Wb = \tilde W  } \\ + \frac{\bar A}{2}\bigg( \frac{(-1)^{i-j}}{\omega-(\varepsilon_b -\bar \omega_p) -i\eta } \\ +\frac{1}{\omega-(\varepsilon_{ab} +\bar \omega_p) +i\eta } \bigg)\,.
\end{multline}

\subsubsection{Second-order expansions}
\begin{itemize}
    \item The standard second-order diagram

The standard second-order self-energy expression (the fifth second-order diagram in Fig.~\ref{Feynman diag o2 self-energy})  is
\begin{multline}
\Sigma^{ \text{stand}\, (2) }_{\mathrm{xc} \,ij\sigma}(\omega)=  -\sum_{k,l=1,2}\int d\omega_1d\omega_2 \,  \Gb_{il\sigma}(\omega-\omega_1) \Wb_{ik}(\omega_1) \\ \times \Gb_{lk\sigma}(\omega-\omega_1-\omega_2)  \Wb_{lj}(\omega_2)\\ \times  \Gb_{kj\sigma}(\omega-\omega_2)e^{-i\omega_1\eta} e^{-i\omega_2\eta} \,.
\end{multline}
This diagram exists for the three different choices of $\Wb$. For the dimer it becomes

\begin{multline}
 \Sigma^{\text{stand} \, (2) }_{\mathrm{xc} \,ij\sigma}(\omega) = \bar \alpha_1 \bigg( \frac{(-1)^{i-j} }{\omega-(\varepsilon_b-2t) -i\eta} \\ + \frac{1}{\omega-(  \varepsilon_{ab}+2t) +i\eta} \bigg)\\ + \bar \alpha_2 \bigg( \frac{(-1)^{i-j} }{\omega-(\varepsilon_b-\bar \omega_p) -i\eta} + \frac{1}{\omega-(\varepsilon_{ab}+\bar \omega_p) + i\eta} \bigg) \\
 + \bar \alpha_3 \bigg( \frac{(-1)^{i-j} }{(\omega-(\varepsilon_b-\bar \omega_p) -i\eta)^2}  \\- \frac{1}{( \omega-(\varepsilon_{ab}+\bar \omega_p) + i\eta)^2} \bigg) \\ + \bar \alpha_4 \bigg( \frac{ 1 }{\omega-(\varepsilon_b-2\bar \omega_p) -i\eta} \\ + \frac{(-1)^{i-j}}{\omega-(\varepsilon_{ab}+2\bar \omega_p) + i\eta} \bigg) \,,
\end{multline}
where
\begin{multline}
\bar \alpha_1 =- \frac{U^2}{8}
- \frac{U\, \bar A }{2(2t-\bar \omega_p)}
+ \frac{U\,\bar A }{2(2t +\bar  \omega_p)}\\
- \frac{\bar A^2}{2} \bigg( \frac{1}{2t+\bar \omega_p} - \frac{1}{2t-\bar\omega_p} \bigg)^2  \underbrace{+ \frac{X^2}{2} }_{\text{only when}\, \Wb = \tilde W}\,,
\end{multline}

\begin{multline}
\bar \alpha_2 = 
\frac{U\, \bar A }{2( 2t+\bar \omega_p)}
+ \frac{U\,\bar A }{2(2t-\bar\omega_p)} 
+ \frac{\bar A^2}{2(2t-\bar\omega_p)^2} \\ - \frac{\bar A^2}{(2t+\bar \omega_p) (2t-\omega_p)} \underbrace{+ \frac{X \bar A}{2t-\bar\omega_p}}_{\text{only when}\,  \Wb = \tilde W  }\,,
\end{multline}

\begin{align}
\bar \alpha_3 = -\frac{\bar A^2}{2(2t-\bar \omega_p)}\,,
\end{align}
and,
\begin{align}
\bar \alpha_4 = \frac{\bar A^2}{2(2t+\bar \omega_p)^2}\,.
\end{align}

\item The self-consistency diagrams\\

The self-consistency diagrams are the first two second-order ones in Fig.~\ref{Feynman diag o2 self-energy}.
\begin{multline}
 \Sigma^{\text{sc} \, (2)}_{\mathrm{xc} \,ij \sigma}(\omega) = i \sum_{k,l=1,2} \int d\omega_1 \Gb_{ik\sigma}(\omega+\omega_1) \bigg( \Sigma^{(1)}_{\mathrm{xc}\,kl\sigma}(\omega+\omega_1) \\ -\underbrace{v^{(1)}_{\mathrm{xc} \, kk} \delta_{kl} }_{ \text{only when} \Wb = \tilde W} \bigg) \Gb_{lj\sigma}(\omega+\omega_1) \Wb_{ji}(\omega_1) e^{i\omega_1\eta}\,.
\end{multline}
This diagram also exists for the three different choices of $\Wb$. Its solution is

\begin{multline}
\Sigma^{\text{sc} \, (2)}_{\mathrm{xc} \,ij\sigma}(\omega)=   \bar \beta_1 \bigg(  \frac{(-1)^{i-j}}{\big(\omega-(\varepsilon_b-\bar \omega_p)-i\eta\big)^2}  \\ +  \frac{1}{\big(\omega-(\varepsilon_{ab}+\bar \omega_p)+i\eta\big)^2} \bigg)
 \\ + \bar \beta_2 \bigg(  \frac{(-1)^{i-j}}{\big(\omega-(\varepsilon_b-\bar \omega_p)-i\eta\big)^2}  -  \frac{1}{\big(\omega-(\varepsilon_{ab}+\bar \omega_p)+i\eta\big)^2} \bigg)\\
 - \frac{\bar A^2}{ 2(2t+\bar \omega_p)^2 } \bigg( \frac{(-1)^{i-j}}{\omega-(\varepsilon_b-\bar \omega_p)-i\eta}  + \frac{1}{\omega-(\varepsilon_{ab}+\bar \omega_p)+i\eta}\bigg)\\
 + \frac{\bar A^2}{2(2t+\bar \omega_p)^2} \bigg( \frac{(-1)^{i-j}}{\omega-(\varepsilon_{ab}+2\bar \omega_p)+i\eta}\\  + \frac{1}{\omega-(\varepsilon_b-2\bar \omega_p)-i\eta}  \bigg)\,,
\end{multline}
where for $\Wb = W$ and $\Wrpa$, $\bar \beta_1 = - \frac{U \bar A}{4} $ and $\bar \beta_2 = - \frac{\bar A^2 }{2(2t + \bar \omega_p)}$. For $\Wb = \tilde W$, 
$\bar \beta_1 = \tilde \beta_1 = - \frac{(U+2 v^{(1)}_{xc} )\tilde A}{4} $ and $ \bar \beta_2 = \tilde \beta_2 = - \frac{\tilde A^2 }{2(2t + \tilde \omega_p)}  + \frac{\tilde AX}{2} $.

\item The $\fxc^{(1)}$ contributions\\

These contributions, as explained in the present paper, appear due to the choice of $\Wb = \tilde W$. Thus, the self-energy at the second-order has additional contributions coming from the $\fxc^{(1)}$, given in Eq.~\eqref{fxc mb o1} and Eq.~\eqref{fxc gaps o1}. The general formula of the self-energy is given in Eq.~\eqref{eq:S-KS}. The first two contributions of Eq.~\eqref{fxc gaps o1} plugged into Eq.~\eqref{eq:S-KS}\, where $\Gb = \Gks$
and $\Wb = \tilde W$ leads to a second-order self-energy written in the dimer basis as
\begin{multline}
 \Sigma^{(2) \, \fxc^{g(1)} }_{\mathrm{xc} \,ij\sigma}(\omega) = - \!\!\!\!\!\!\sum_{k,l,m,n=1,2} \int d\omega_1 d\omega_2 \Gb_{ij\sigma}(\omega_1+\omega) \tilde{W}_{li}(\omega_1) \\\times \chi_{0\,jk\sigma}^{-1}(\omega_1) \Gb_{kl\sigma}(\omega_1+\omega_2) \Gb_{lm\sigma}(\omega_2) \bigg(  \Sigma^{(1)}_{\text{xc} \, mn\sigma}(\omega_2) \\ -v^{(1)}_{\text{xc}\, mm} \delta_{mn} \bigg)  \Gb_{nk\sigma}(\omega_2) e^{i\omega_1\eta}e^{2i\omega_2\eta}\,.
\end{multline}
Its analytical solution is

\begin{multline}
 \Sigma^{(2) \, \fxc^{g(1)}}_{\mathrm{xc}\,ij\sigma}(\omega) = (-1)^{i-j}\lambda_0 + (-1)^{i-j }\frac{\lambda_1}{\omega-(\ehks-2t)-i\eta} \\ + \frac{\lambda_2}{\omega-(\elks+2t)+i\eta} + (-1)^{i-j }\frac{\lambda_3}{\omega-(\ehks-\tilde \omega_p)-i\eta} \\ + \frac{\lambda_4}{\omega-(\elks+\tilde \omega_p)+i\eta}\,,
\end{multline}
where

\begin{multline}
 \lambda_0 = - \frac{X(U+2X)}{16t} - \frac{X\tilde{A}}{2(2t+\tilde\omega_p)^2} + \frac{(U+2X)\tilde{A}}{8t(2t+\tilde \omega_p)} \,,
\end{multline}

\begin{multline}
 \lambda_1 = - \frac{U(U+2X)}{8}  + \frac{U\tilde{A}}{4(2t+\tilde \omega_p)} - \frac{U\tilde{A}}{4(2t- \tilde \omega_p)} \\ - \frac{(U+2X)\tilde{A}}{4(2t+\tilde\omega_p)}  + \frac{\tilde{A}^2}{2(2t+\tilde\omega_p)^2} - \frac{\tilde{A}^2}{2(2t+\tilde\omega_p)(2t-\tilde\omega_p)} \\
+ v^{(1)}_{xc}\bigg(  - \frac{U+2X}{4} + \frac{\tilde{A}}{2(2t+\tilde\omega_p)} - \frac{\tilde{A}}{2(2t-\tilde \omega_p)} \bigg) \,,
\end{multline}

\begin{multline}
 \lambda_2 =  \frac{(U+2X)^2}{8} - \frac{(U+2X)\tilde{A}}{4(2t+\tilde\omega_p)} + \frac{(U+2X)\tilde{A}}{4(2t-\tilde\omega_p)} \\ - \frac{(U+2X)\tilde{A} }{4(2t+\tilde\omega_p)} + \frac{\tilde{A}^2}{2(2t+\tilde\omega_p)^2} - \frac{\tilde{A}^2}{2(2t-\tilde\omega_p)(2t+\tilde\omega_p)} \\
 + v_{xc}^{(1)} \bigg( \frac{U+2X}{4} - \frac{\tilde{A}}{2(2t+\tilde\omega_p)} + \frac{\tilde{A}}{2(2t-\tilde\omega_p)} \bigg)\,,
\end{multline}

\begin{multline}
 \lambda_3 = \frac{(U+2X)\tilde{A}}{16t} - \frac{(U+2X)\tilde{A}}{4(2t+\tilde \omega_p)} - \frac{U\tilde{A}}{16t} + \frac{U\tilde{A}}{4(2t-\tilde\omega_p)}  \\ + \frac{\tilde{A}^2}{2(2t+\tilde\omega_p)(2t-\tilde\omega_p)} + \frac{\tilde{A}^2(2t-\tilde\omega_p)}{ 8t(2t+\tilde\omega_p)^2} - \frac{\tilde{A}^2}{8t(2t+\tilde\omega_p)} \\+ \frac{\tilde{A}^2}{2(2t+\tilde\omega_p)^2}
 + v_{xc}^{(1)} \bigg(   \frac{\tilde{A}}{2(2t-\tilde\omega_p)} - \frac{\tilde{A}}{2(2t+\tilde\omega_p)} \bigg)\,,
\end{multline}

and
\begin{multline}
\lambda_4 =  \frac{(U+2X)\tilde{A}}{16t} - \frac{(U+2X)\tilde{A}}{4(2t-\tilde\omega_p)} - \frac{U\tilde{A}}{16t} + \frac{U\tilde{A}}{4(2t+\tilde\omega_p)}\\ + \frac{\tilde{A}^2(2t-\tilde\omega_p)}{8t(2t+\tilde\omega_p)^2} - \frac{\tilde{A}^2}{8t(2t+\tilde\omega_p)} + \frac{\tilde{A}^2}{2(2t+\tilde\omega_p)^2}  \\+  \frac{\tilde{A}^2}{2(2t-\tilde\omega_p)(2t+\tilde\omega_p)} 
+ v_{xc}^{(1)} \bigg(  - \frac{\tilde{A}}{2(2t-\tilde\omega_p)}\\ + \frac{\tilde{A}}{2(2t+\tilde\omega_p)} \bigg)\,.
\end{multline}

The second two contributions of the gap term of $\fxc^{(1)}$ in Eq.~\eqref{fxc gaps o1} plugged into Eq.~\eqref{eq:S-KS} leads to the following second-order self-energy
\begin{multline}
\Sigma^{(2) \, \fxc^{g'(1)}}_{\mathrm{xc} \,ij\sigma}(\omega) = - \sum_{k,l,m,n=1,2} \int d\omega_2 d\omega_4\, \Gb_{ij\sigma}(\omega_2+\omega) \\ \times \tilde{W}_{ki}(\omega_2)  \chi^{-1}_{0\, jl\sigma}(\omega_2) \Gb_{kl\sigma}(\omega_4) \Gb_{lm\sigma}(\omega_2+\omega_4) \\ \times \bigg( \Sigma^{(1)}_{\mathrm{xc}\, mn\sigma}(\omega_2+\omega_4)  - v^{(1)}_{\text{xc}\, mm} \delta_{mn} \bigg) \Gb_{nk\sigma}(\omega_2+\omega_4) \\\times e^{2i\omega_4\eta} e^{i\omega_2\eta}\, 
\end{multline}
with $\Gb = \Gks$.

\begin{multline}
 \Sigma^{(2) \, \fxc^{g'(1)}}_{\mathrm{xc} \, ij\sigma}(\omega) = (-1)^{i-j}\kappa_0 + (-1)^{i-j }\frac{\kappa_1}{\omega-(\ehks-2t)-i\eta} \\ + \frac{\kappa_2}{\omega-(\elks+2t)+i\eta} + (-1)^{i-j }\frac{\kappa_3}{\omega-(\ehks-\tilde \omega_p)-i\eta} \\ + \frac{\kappa_4}{\omega-(\elks+\tilde \omega_p)+i\eta}\,,
\end{multline}
where

\begin{equation}
 \kappa_0 = - \frac{X(U+2X)}{16t} - \frac{X\tilde A}{2(2t+\tilde\omega_p)^2} + \frac{(U+2X)\tilde A}{8t(2t+\tilde \omega_p)} \,,
\end{equation}

\begin{multline}
 \kappa_1 = \frac{(U+2X)(U+2X)}{8}  - \frac{(U+2X)\tilde A}{4(2t+\tilde \omega_p)} + \frac{(U+2X)\tilde A}{4(2t-\tilde \omega_p)} \\ - \frac{(U+2X)\tilde A}{4(2t+\tilde \omega_p)} + \frac{\tilde A^2}{2(2t+\tilde \omega_p)^2} - \frac{\tilde A^2}{2(2t+\tilde \omega_p)(2t-\tilde \omega_p)} \\
+ v^{(1)}_{xc}\bigg(   \frac{U+2X}{4} - \frac{\tilde A}{2(2t+\tilde \omega_p)} + \frac{\tilde A}{2(2t-\tilde \omega_p)} \bigg) \,,
\end{multline}

\begin{multline}
 \kappa_2 =  - \frac{U(U+2X)}{8} + \frac{U\tilde A}{4(2t+\tilde \omega_p)} - \frac{U\tilde A}{4(2t-\tilde \omega_p)} \\ - \frac{(U+2X)\tilde A}{4(2t+\tilde \omega_p)} + \frac{\tilde A^2}{2(2t+\tilde \omega_p)^2} - \frac{\tilde A^2}{2(2t-\tilde \omega_p)(2t+\tilde \omega_p)} \\
 + v_{xc}^{(1)} \bigg(- \frac{U+2X}{4} + \frac{\tilde A}{2(2t+\tilde \omega_p)} - \frac{\tilde A}{2(2t-\tilde \omega_p)} \bigg)\,,
\end{multline}

\begin{multline}
 \kappa_3 = \frac{(U+2X)\tilde A}{16t} - \frac{(U+2X)\tilde A}{4(2t-\tilde \omega_p)} - \frac{U\tilde A}{16t} + \frac{U\tilde A}{4(2t+\tilde \omega_p)} \\ + \frac{\tilde A^2}{2(2t+\tilde \omega_p)(2t-\tilde \omega_p)} + \frac{\tilde A^2(2t-\tilde \omega_p)}{ 8t(2t+\tilde \omega_p)^2} - \frac{\tilde A^2}{8t(2t+\tilde \omega_p)} \\+ \frac{\tilde A^2}{2(2t+\tilde \omega_p)^2}
 + v_{xc}^{(1)} \bigg(   -\frac{\tilde A}{2(2t-\tilde \omega_p)}\\ + \frac{A}{2(2t+\tilde \omega_p)} \bigg)\,,
\end{multline}

and,
\begin{multline}
\kappa_4 =  \frac{(U+2X)\tilde A}{16t} - \frac{(U+2X)\tilde A}{4(2t+\tilde \omega_p)} - \frac{UA}{16t} + \frac{UA}{4(2t-\tilde \omega_p)} \\+ \frac{\tilde A^2(2t- \tilde  \omega_p)}{8t(2t+\tilde \omega_p)^2} - \frac{\tilde A^2}{8t(2t+\tilde \omega_p)} + \frac{\tilde A^2}{2(2t+\tilde \omega_p)^2} \\ +  \frac{\tilde A^2}{2(2t-\tilde \omega_p)(2t+\tilde \omega_p)} 
+ v_{xc}^{(1)} \bigg(   \frac{\tilde A}{2(2t-\tilde \omega_p)} \\- \frac{\tilde A}{2(2t+\tilde \omega_p)} \bigg)\,.
\end{multline}

Finally, the many-body contribution part of $\fxc^{(1)}$ (Eq.~\eqref{fxc mb o1}) yields the following second-order contribution

\begin{multline}
 \Sigma^{(2)\, \fxc^{\text{mb}(1)}}_{\mathrm{xc} \,pn\sigma}(\omega) = -i \sum_{i,j,k,l\,\sigma'} \int d\omega_1 d\omega_2 d\omega' \, \Gb_{pn\sigma}(\omega_2+\omega)\\ \times \chi^{-1}_{0\,nj}(\omega_2) \tilde{W}_{ik}(\omega_1) \Gb_{ji\sigma'}(\omega') \Gb_{kj\sigma'}(\omega' - \omega_2) G_{il\sigma'}(\omega' - \omega_1) \\ \Gb_{lk\sigma'}(\omega' - \omega_1 - \omega_2) \tilde{W}_{lp}(\omega_2)e^{- i\omega_1\eta}e^{- i\omega_2 \eta}e^{2i\omega' \eta}\,,
\end{multline}
where $\Gb = \Gks$. It reads

\begin{multline}
\Sigma^{(2)\, \fxc^{\text{mb}(1)} }_{\mathrm{xc}\,ij \sigma}(\omega) = (-1)^{i-j} \bigg( \frac{X(U+2X)}{8t} - (\frac{U}{2} + X ) \frac{4\tilde A}{8t(2t+\tilde \omega_p)} \\  - \frac{U\tilde A}{2(2t+\tilde \omega_p)^2}\bigg)  + \bigg( \frac{U(U+2X)}{8} -\frac{U\tilde A}{4(2t+\tilde \omega_p)} \\+ \frac{U\tilde A}{4(2t-\tilde \omega_p)} \bigg) \bigg(  \frac{1}{\omega-(\elks + 2t) + i\eta} +  \frac{(-1)^{i-j}}{\omega-(\ehks - 2t) - i\eta} \bigg) \\
+ \bigg( \frac{U\tilde A}{8t} - \frac{U\tilde A}{4(2t-\tilde \omega_p)} - \frac{U\tilde A}{4(2t+\tilde \omega_p)} \\+ \frac{\tilde A}{8t} \big(  \frac{4\tilde A}{2t+\tilde \omega_p} -(U+2X)  \big) \bigg) \bigg(  \frac{1}{\omega-(\elks + \tilde \omega_p) + i\eta} \\+  \frac{(-1)^{i-j}}{\omega-(\ehks -\tilde  \omega_p) - i\eta} \bigg)\,.
\end{multline}

Note that the three diagrams that stem from $\fxc^{(1)}$ depend on $\chi_0^{-1}$, which, in the symmetric Hubbard dimer, has an arbitratry bonding-bonding term (i.e., $\chi_0$ is not invertible). The same holds for the exact $\chi$. However, as already explained in \cite{elsahili2023total}, this does not affect the results. The $\chi_0$ matrix, in the $ij$ basis of the dimer, is
\begin{equation}
    \chi_0(\omega) = \begin{pmatrix}
        \chi_{0\,11 \uparrow}(\omega) & \chi_{0\,12 \uparrow}(\omega) & 0 & 0\\
        \chi_{0\,21 \uparrow}(\omega)  & \chi_{0\,22 \uparrow}(\omega) & 0 & 0\\
        0 & 0 & \chi_{0\,11 \downarrow}(\omega) & \chi_{0\,12 \downarrow}(\omega)\\
        0 & 0 & \chi_{0\,21 \downarrow}(\omega) & \chi_{0\,22 \downarrow}(\omega)
    \end{pmatrix}\,,
\end{equation}
where $\chi_{0\,ij\uparrow}(\omega) = \chi_{0\,ij\downarrow}(\omega) = \frac{(-1)^{i-j}}{4} \bigg( \frac{1}{\omega-2t+i\eta} - \frac{1}{\omega+2t-i\eta} \bigg)$\,. As one can see, the problem is that $\chi_{0\,12} = - \chi_{011}$. 
However, in the equations only the sum $\chi^{-1}_{0\,11} - \chi^{-1}_{0\,12}$ is needed, and the problem drops out.

\end{itemize}

\subsubsection{Exchange-correlation energy}

To calculate the $\Exc$ contribution to $\E$, we use the last term of Eq.~\eqref{GM in self-energy}. In the Hubbard dimer $ij$ basis it reads
\begin{equation}
    \Exc = - \frac{i}{2} \sum_{i,j=1,2, \, \sigma=\uparrow,\downarrow} \int d\omega \, \Gb_{ij\sigma}(\omega) \Sigma_{\text{xc} \, ji \sigma}(\omega) e^{2i\omega\eta}\,.
    \label{exc HD formula}
\end{equation}
We provide analytical solutions for the exact and approximate $\Exc$.

\begin{itemize}
  \item Exact solution:  the exact $\Exc$ can be obtained in many different ways such as using the exact GF and $\Sigmaxc$ in Eq.~\eqref{exc HD formula}. However, here we obtained it using the exact response function $\chi$, where $\Exc$ can be directly expressed in terms of $\chi$ as shown in \cite{elsahili2023total}. Thus, 
  
  \begin{equation}
      \Exc^{\text{exact}} = - U + \frac{2U}{a^2}\,.
  \end{equation}
    
    \item First-order expansion:

    \begin{equation}
        \Exc^{(1)} = - \frac{U}{2} + \underbrace{X}_{\text{only when $\Wb = \tilde W$}} - \frac{2\bar A}{2t+ \bar\omega_p}\,.
        \label{o1 xc energy}
    \end{equation}
One can already see that $\Exc$ at \firstord{} in $\tilde W$ is exact. We write down, below, the proof
\begin{align}
    \Exc^{(1)} & =  - \frac{U}{2} + X - \frac{2\tilde A}{2t+ \tilde \omega_p}\\
    &= - U + \frac{2U\tilde \omega_p}{2ta^2} - \frac{2U(2t-\tilde \omega_p)}{2ta^2}\\
    &= - U  + \frac{2U}{a^2}\,.
\end{align}

\item Second-order expansion: the second-order expansions of the $\Exc$ in the different choices of $\Wb$ with their correspondant choices of GFs are shown in Section.~\ref{detailed deriv}.

\begin{multline}
    \Exc^{(2)\, \text{stand}} =  \frac{U^2}{8t} - \frac{U\bar A}{t(2t+\bar  \omega_p)} - \frac{2U\bar A}{(2t+\bar  \omega_p)^2} \\ + \frac{4\bar{A}^2}{2t(2t+\bar\omega_p)^2}
    \underbrace{- \frac{X^2}{2t}  - \frac{4X\tilde  A}{(2t-\tilde \omega_p)(2t+\tilde  \omega_p)} }_{\text{only when}\, \Wb = \tilde{W} } \,.
\end{multline}

\begin{equation}
    \Exc^{(2)\, \text{sc}}= \Exc^{G^{(1)}\Sigmaxc^{(1)} } = \frac{4 \bar A^2}{(2t+\bar  \omega_p)^3} - \underbrace{ \frac{2 \tilde AX}{(2t+\tilde  \omega_p)^2}}_{ \text{only when}\, \Wb = \tilde{W}   } \,.
\end{equation}
The additional diagrams when $\Wb = \tilde W$ yields
\begin{multline}
    \Exc^{(2)\, \fxc^{g(1)} }=  \Exc^{(2)\, \fxc^{g'(1)} } = - \frac{4 \tilde A^2}{(2t+\tilde \omega_p)^3}\\ + \frac{2 \tilde AX}{(2t+\tilde \omega_p)^2} \,.
\end{multline}

\begin{multline}
    \Exc^{(2)\, \fxc^{\text{mb}(1) } } =  -\frac{U^2}{8t} + \frac{U\tilde A}{t(2t+\tilde \omega_p)} + \frac{2U\tilde A}{(2t+\tilde \omega_p)^2}\\ - \frac{4\tilde{A}^2}{2t(2t+\tilde{\omega}_p)^2} + \frac{X^2}{2t} + \frac{4X\tilde A}{(2t-\tilde \omega_p)(2t+\tilde \omega_p)} \,.
\end{multline}
We define 
\begin{multline}
    \Exc^{(2)} = \Exc^{(1)} +   \Exc^{(2)\,\text{stand}} +  \Exc^{(2)\, \text{sc}} + \Exc^{G^{(1)}\Sigmaxc^{(1)} }\\ + \underbrace{  \Exc^{(2)\, \fxc^{g(1)} } +  \Exc^{(2)\, \fxc^{g'(1)} } +   \Exc^{(2)\, \fxc^{\text{mb}(1) } }}_{ \Wb = \tilde W}\,,
    \label{tot o2 xc energy}
\end{multline}
where we recall that the contributions underbraced appear when $\Wb = \tilde W$, which, in this case, cancel out with the other contributions leading to $\Exc = \Exc^{(1)} = \Exc^{\text{exact}}$. 
\end{itemize}

\subsubsection{Kinetic energy}

The kinetic energy integral in the dimer is
\begin{equation}
 \Ek = \frac{it}{2\pi} \sum_{i,j=1,2,i\neq j\,, \sigma=\uparrow,\downarrow} \int d\omega \,\Gb_{ij \, \sigma}(\omega) e^{i\omega\eta}\,.   
\label{ek integral dimer}
\end{equation}
Here, $\Gb$ can be solution of the FDE or obtained by EDE. The integral in \eqref{ek integral dimer} is solved numerically.

Because of the symmetry constraints of our model, the natural orbitals are the bonding and anti-bonding orbitals. It is therefore useful to write \eqref{ek integral dimer} in that basis,
\begin{eqnarray}
 \Ek &=& \frac{it}{2\pi} \int d\omega \Big (\Gb_{bb }(\omega)-\Gb_{abab}(\omega)\Big ) e^{i\omega\eta}\nonumber\\
 &=& t \Big ( f_b-f_{ab}\Big )\,,
\label{ek integral dimer-ba}
\end{eqnarray}
where the factor 2 due to the sum over spins is canceled by a factor $1/2$ due to the basis transformation. The numbers $f_b$ and $f_{ab}$ are the two eigenvalues of the one-body reduced density matrix, the occupation numbers of the natural orbitals. Including the sum over spins, their difference equals 2 in the non-interacting limit, where all the weight is in the occupied quasi-particle peaks, and it decreases with increasing correlation strength, where weight is shifted from the quasi-particles to the satellites. Note that in the limit of $t\to 0$ one may obtain the correct vanishing kinetic energy even with wrong occupation numbers, because of the prefactor $t$. \\

\subsubsection{Quasi-particle energies and satellites \label{qp calc details}}
\begin{figure*}
    \centering
    \includegraphics[width=0.9\linewidth]{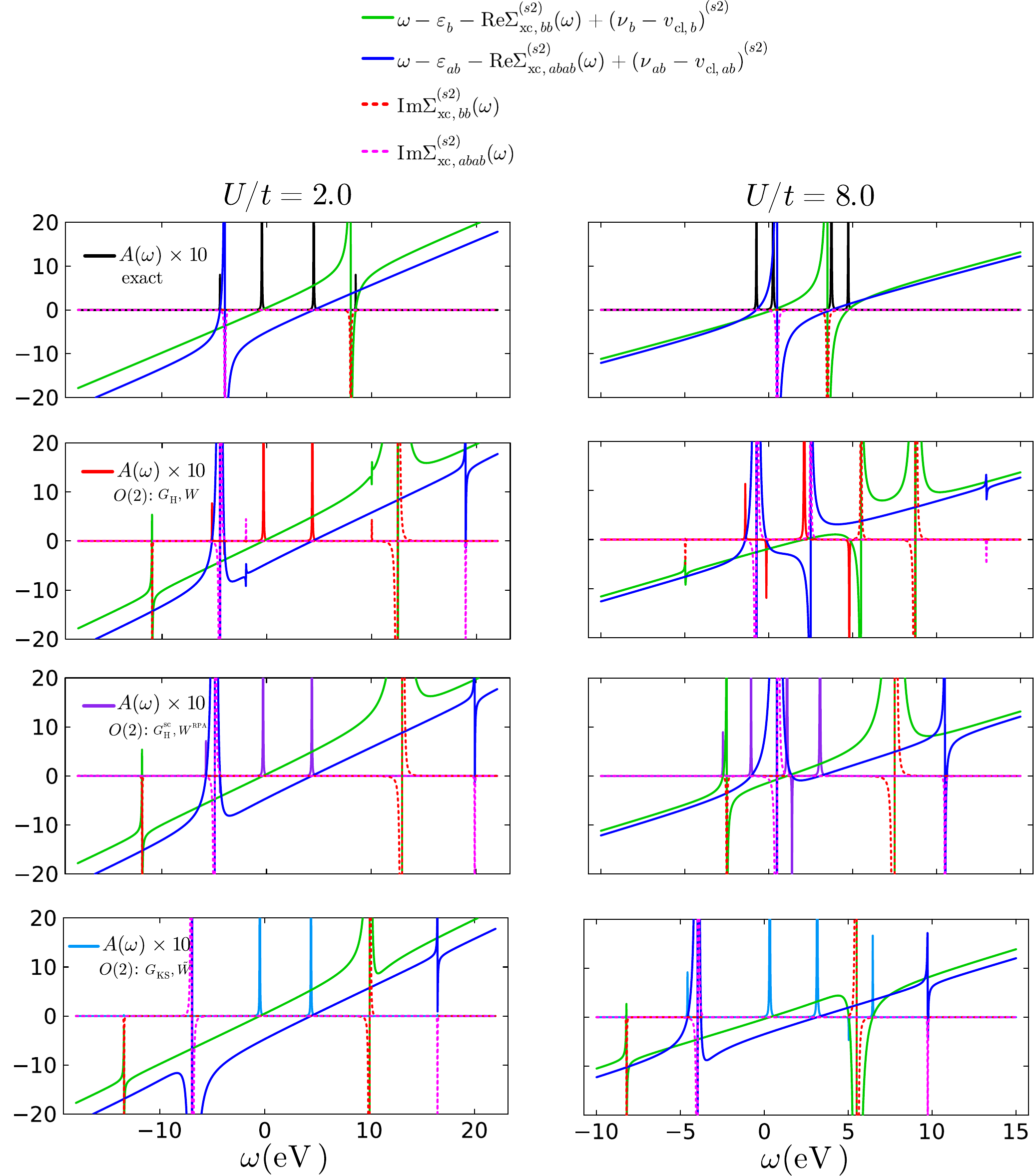}
    \caption{Spectral features obtained from exact solutions are compared with those from second-order approximations. The exact results are shown in the top panel, while the second-order results—$(\GH, W)$, $(\GH^{\text{sc}}, \Wrpa)$, and $(\Gks, \tilde{W})$—are shown in the second, third, and fourth panels, respectively. Two representative values of $U/t$ are considered to highlight the evolution of the spectra. The spectral functions $A(\omega)$ in all panels are the exact (black), $(\GH, W)$ (red), $(\GH^{\text{sc}}, W^{\rm RPA})$ (violet), and $(\Gks, \tilde{W})$ (light blue). The shifted real parts of the self-energy for the bonding–bonding (bb) and antibonding–antibonding (abab) contributions are represented by solid green and blue curves, respectively. The corresponding imaginary parts of the self-energies are shown as dotted red (bb) and magenta (abab) curves. For the exact solution, the exact $\Sigmaxc$ is used, which, in this figure, we calculate from $\Sigmaxc = G_0^{-1} - \vcl - G^{-1}$, with $G$ the exact Green's function.}
    \label{spectral features exact vs o2}
\end{figure*}

The quasi-particle (QP) energies $\epsilon_b$ and $\epsilon_{{ab}}$ shown in Fig.~\ref{qp energies plot} are obtained by solving the QP equations (second choice in \eqref{eq:qp-choices}) in the bonding (b)/anti-bonding (ab) basis, where all one-body quantities are diagonal. The exact QP energies are
$\epsilon^{\text{exact}}_b = \epsilon_0 +t - (c-U)/2$ and $\epsilon^{\text{exact}}_{ab} = \epsilon_0 -t + (c+U)/2$.  For certain approximations and parameter ranges, there is no QP solution. The reason is the wrong analytical behaviour of the imaginary part of the self-energy. This can be appreciated in Fig.~\ref{spectral features exact vs o2}.  The problematic features, for example, double poles, become so strong for small $t$ that they dominate and prevent the zero-crossing of the blue lines related to the real part. This does not happen when $\Wb=\tilde W$.

Satellites are obtained by solving the Dyson equation and calculating the spectral function from the imaginary part of the resulting Green's function.

\bibliography{biblio}

\end{document}